\documentclass[12pt]{article}
\pdfoutput=1
\usepackage{graphicx,epstopdf,amssymb,amsfonts,amsmath,amsthm,array,
mathrsfs,amscd}
\usepackage{todonotes} 
\newcommand{\pbs}[1]{\let\temp=\\#1\let\\=\temp}
\numberwithin{equation}{section}
\def\be{\begin{equation}}\def\ee{\end{equation}}
\def\cvp{\raise 2pt\hbox{,}}

 \def\d{{\rm d}}

\def\diagramb{\begin{picture}(6,6)(0,0)
\put(3,2){\circle{6}}
\put(3,-1){\line(0,1){6}}
\end{picture}}
\def\diagrambbig{\begin{picture}(10,10)(0,0)
\put(5,3.5){\circle{10}}
\put(5,-1.5){\line(0,1){10}}
\end{picture}}
\def\diagramc{\begin{picture}(12,6)(0,0)
\put(3,2){\circle{6}}\put(9,2){\circle{6}}
\end{picture}}
\def\diagramcbig{\begin{picture}(20,10)(0,0)
\put(5,3.5){\circle{10}}\put(15,3.5){\circle{10}}
\end{picture}}
\def\diagramd{\begin{picture}(15,6)(0,0)
\put(3,2){\circle{6}}\put(6,2){\line(1,0){3}}\put(12,2){\circle{6}}
\end{picture}}
\def\diagramdbig{\begin{picture}(26,10)(0,0)
\put(5,3.5){\circle{10}}\put(10,3.5){\line(1,0){5}}\put(20,3.5){\circle{10}}
\end{picture}}

\def\d{\partial}

\def\d{{\rm d}}

\def\Det{{\rm Det}}

\def\a{\alpha}
\def\b{\beta}
\def\g{\gamma}
\def\G{\Gamma}
\def\dd{\delta}

\def\m{\mu}

\def\k{\kappa}
\def\l{\lambda}

\def\s{\sigma}
\def\f{\phi}
\def\p{\psi}

\def\D{\Delta}
\def\z{\zeta}

\def\vf{\varphi}
\def\L{\Lambda}
\def\o{\omega}

\def\wh{\widehat}
\def\wt{\widetilde}

\def\l{\lambda}

\def\del{\partial}

\def\ba{\begin{eqnarray}}
\def\ea{\end{eqnarray}}

\theoremstyle{plain}

\theoremstyle{definition}
\theoremstyle{remark}

\def\imath#1#2#3{{\it Invent math }{\bf #1} (#2) #3}

\def\Gd{{\cal G}^{(2)}}

\begin{document}
%
%


{\pagestyle{empty}
\parskip 0in
\

\vfill
\begin{center}

{\Large \bf 2D quantum gravity on compact Riemann surfaces}

\medskip

{\Large \bf and two-loop partition function:}

\medskip

{\Large \bf circumventing the $c=1$ barrier?}
\vspace{0.4in}

Adel B{\scshape ilal} and Laetitia L{\scshape educ}
\\
\medskip
\it {Centre National de la Recherche Scientifique\\
Laboratoire de Physique Th\'eorique de l'\'Ecole Normale Sup\'erieure\\
24 rue Lhomond, F-75231 Paris Cedex 05, France}

\smallskip

{\tt adel.bilal@lpt.ens.fr, laetitia.leduc@lpt.ens.fr}
\end{center}
\vfill\noindent
We study two-dimensional quantum gravity on arbitrary genus Riemann surfaces in the K\"ahler formalism where the basic quantum field is the (Laplacian of the) K\"ahler potential. We do a careful first-principles computation of the fixed-area partition function $Z[A]$ up to and including all two-loop contributions. This includes genuine two-loop diagrams as determined by the Liouville action, one-loop diagrams resulting from the non-trivial measure on the space of metrics, as well as one-loop diagrams involving various counterterm vertices. Contrary to what is often believed, several such counterterms, in addition to the usual cosmological constant, do and must occur. We consistently determine the relevant counterterms from a one-loop computation of the full two-point Green's function of the K\"ahler field. Throughout this paper we  use the general spectral cutoff regularization developed recently and which is well-suited for multi-loop computations on curved manifolds. At two loops, while all  ``unwanted" contributions to $\ln (Z[A]/Z[A_0])$ correctly cancel, it appears that the {\it finite} coefficient of $\ln (A/A_0)$ does depend on the finite parts of certain counterterm coefficients, i.e. on the finite renormalization conditions one has to impose. There exists a choice that reproduces the famous KPZ-scaling, but it seems to be only {\it one} consistent choice among others. Maybe, this hints at the possibility that  other renormalization conditions could eventually provide a way to circumvent the famous $c=1$ barrier.

\vfill
\medskip
%
\begin{flushleft}
\end{flushleft}
\newpage\pagestyle{plain}

}

{\parskip -0.3mm
\small{\tableofcontents}}
\newpage
\setcounter{page}{1}


%
%
\section{Introduction}

\subsection{Introduction and motivation}

Since Polyakov's seminal paper \cite{Liouville1}, conformal matter coupled to quantum gravity on two-dimensional manifolds has been studied intensely, both in the discretized approach \cite{triang} and in the continuum approach \cite{Zamolod, KPZ, DDK, matrixold, matrixmatter}.
Most of the continuum literature uses conformal gauge, where the effect of having integrated out the matter results in an effective gravity action being the Liouville action with a coefficient $\k^2\sim\break -(c-26)$ where $c$ is the matter central charge and the $-26$ accounts for the gauge fixing (ghosts). One of the simplest, yet interesting objects to study in this quantum gravity is the partition function at fixed area $Z[A]$. 

One way to define this partition function $Z[A]$  on a Riemann surface of genus $h$, with metric $g$ of area $A$, is to choose the conformal gauge with a background metric $g_0$ and a conformal factor $\s$ such that $g=e^{2\s}g_0$. Then $Z[A]$ can be formally written as
\be\label{ZAdef}
Z[A]=\int {\cal D}\s \exp\left( -\frac{\k^2}{8\pi} S_L-S_{\rm cosm}\right) \, 
\dd \left(A-\int \d^2 x \sqrt{g_0}\, e^{2\s}\right)
 \ ,
\ee
where the Liouville action and $\k^2$ are given by
\be\label{Liouvaction}
S_L[\s]=\int\d^2 x \sqrt{g_0} \, \big( \s \D_0\s + R_0 \s\big)
\quad , \quad \k^2=\frac{26-c}{3} \ ,
\ee
while the cosmological constant term simply is $S_{\rm cosm}= \m_c^2 \int\d^2 x \sqrt{g_0} \, e^{2\s} = \m_c^2 A$, 
and the delta-function restricts the integration to metrics of area $A$. The mesure ${\cal D}\s$ for the conformal factor can be derived from the standard metric on the space of metrics and is a complicated non-flat measure. Many of the difficulties in dealing with this quantum gravity theory originate from this  measure being non-trivial.

KPZ studied two dimensional gravity for genus zero in the light-cone gauge instead \cite{KPZ}. Using the relation with an ${\rm SL}(2)$ current algebra they derived a remarkable formula relating the scaling dimensions $\D$ of conformal primary operators coupled to gravity and their undressed conformal dimensions $\D^{(0)}$:
\be\label{algKPZ}
\D-\D^{(0)}=\frac{ \big(\sqrt{25-c}-\sqrt{1-c}\,\big)^2}{24} \D (1-\D) \ ,
\ee
known as algebraic KPZ relation. The scaling of the partition function then is obtained from the dressing of the identity operator ($\D_{\rm id}^{(0)}=0$) and leads to a scaling
\be\label{KPZscaling}
Z[A]\sim e^{-\m_c^2 A} A^{\g_{\rm str} -3} \ ,
\ee
with $\g_{\rm str}=\D_{\rm id}$. Eq.\eqref{algKPZ} then yields
$\g_{\rm str}= 2-2 \frac{\sqrt{25-c}}{\sqrt{25-c}-\sqrt{1-c}}$. This formula gives the correct scaling for certain random lattice models corresponding to specific values of~$c$.

On the other hand, working in conformal gauge, and
using several simplifying assumptions together with consistency conditions, ref.~\cite{DDK} have extended these  remarkable formulae to arbitrary genus. In particular they find for the  area dependence of the partition function
\be\label{gammastr}
\g_{\rm str}=2 + 2(h-1) \frac{ \sqrt{25-c}}{\sqrt{25-c}-\sqrt{1-c}} \ .
\ee
More recently, ref. \cite{Duplantier} has given a more rigorous, though more abstract derivation of these formulae for $c\le 1$ (and $h=0$) through a probabilistic reformulation. An alternative, more physical derivation can be found in \cite{Davidrig}.

Probably the most puzzling property of these formulae is that they only seem to work for $c\le 1$, since for $c>1$ (actually $1<c<25$) they turn complex. This is the so-called $c=1$ barrier. 
It has been argued that, for $c>1$, the geometry is dominated by configurations that no longer are smooth and that the surface develops spikes and a fractal character.

It is clearly desirable to do a first-principles quantum field theory computation of $Z[A]$ on a Riemann surface of arbitrary genus, using a well-defined physical regularization scheme and a precise definition of the measure ${\cal D}\s$ as it follows from the usual metric on the space of metrics. This is what has been initiated in \cite{BFK} and what we will continue in this paper. We will compute $Z[A]$ in a loop-expansion where $\frac{1}{\k^2}$ is the loop-counting parameter. The corresponding loop-expansion of \eqref{gammastr} is
\be\label{gammastrexp}
\g_{\rm str}=\frac{1}{2}(h-1)\k^2 +\frac{19-7h}{6} + 2(1-h)\frac{1}{\k^2} 
+ {\cal O}(\k^{-4}) \ .
\ee
To compute the partition function at fixed area it is obviously  convenient to have a parametrization of the metric where the area appears explicitly as a ``coordinate" on the space of metrics. This is naturally implemented in the K\"ahler approach where the metric is determined by  a fixed background metric, the area $A$ and the Laplacian (in the background metric) of the K\"ahler potential. 
In this formalism, the measure on the space of metrics  is given in terms of $\d A$, the standard flat measure on the space of K\"ahler potentials, and various non-trivial determinants. Expanding these determinants and the interactions in the Liouville action in powers of $\frac{1}{\k^2}$ then generates the loop-expansion. 

To be well-defined, of course, we also need to implement a consistent regularization.  Here, as was also done in ref \cite{BFK},  we employ the general spectral cutoff regularization developed in \cite{BF} that is well-suited for use on curved manifolds. It is a generalization of the $\zeta$-function regularization that works  at one loop, to a general regularization scheme that works for multi-loop computations on curved manifolds.
Its basic objects are the heat kernel and generalized heat kernels defined on the manifold for which exist well-known formulae for the asymptotic ``small $t$"  behaviour.

In ref.~\cite{BFK} the partition function $Z[A]$ was computed, with this regularization scheme, up to and including the one-loop contributions, using a more general quantum gravity action that is a sum of the Liouville as well as Mabuchi  \cite{FKZMab} actions. This gave a definite result for $\g_{\rm str}$ which for the pure Liouville gravity reduced to $\g_{\rm str}^{0,1-{\rm loop}}=\frac{1}{2}(h-1)\k^2 +\frac{19-7h}{6}$ in agreement with \eqref{gammastrexp}. Although satisfying, the agreement was, maybe, not too much a surprise. Indeed, the truely non-trivial nature of the determinants coming from the measure over the space of metrics only shows up beyond one loop, starting at two loops. It is thus quite intriguing to try and compute the two-loop contributions to the fixed-area partition function. This is what we will do in the present paper.

Let us mention that the present work grew out of the attempt to compute the two-loop fixed-area partition function $Z[A]$ in another regularization scheme where the infinite-dimensional space of inequivalent metrics is replaced by the finite-dimensional space of inequivalent Bergmann metrics of degree $N$.  As shown in \cite{FKZ}, in the K\"ahler formalism, this  amounts to  expanding a certain function of the K\"ahler potential on a specific finite-dimensional basis $\bar s_\a\, s_\b$ of functions on the Riemann surface. If the Riemann surface is a sphere, the $s_\a$ are the $N+1$  holomorphic sections of the $O(N)$ line bundle. Equivalently, this can be seen as expanding on the truncated $(N+1)^2$-dimensionsl basis of spherical harmonics $Y_l^m$ with $l\le N$. Then, for genus 0, the relevant two-loop diagrams \cite{BFunpub} involve complicated sums over products of Clebsch-Gordan coefficients, truncated in a specific way depending on $N$. In \cite{BFunpub} the large $N$ behaviour of these sums has been evaluated, analytically for some of them and numerically for the others. To be able to read off the value of $\g_{\rm str}$ necessitates to reliably compute the subleading terms in the large-$N$ asymptotics. While the evaluations of \cite{BFunpub} were precise enough to obtain the leading, as well as some subleading asymptotics, the result was not the one expected. In this approach,\break $(N+1)^2\sim A\L^2$ plays the role of a sharp cutoff and, as discussed in \cite{BF}, such sharp cutoffs generally are plagued with difficulties and often do not allow a well-defined large-$N$ asymptotic expansion beyond the leading term. This was one of our motivations when developing instead the general spectral cutoff regularization that does have a well-defined large-$\L$ asymptotic expansion. It was used in \cite{BFK} to compute $Z[A]$ up to one loop and it is used in the present paper to do the two-loop computation.

\subsection{Outline and summary of the results}

As it turns out, this two-loop computation actually is quite an enterprise. While the non-trivial two-loop contribution from the measure determinant is  rather easy to work out in this K\"ahler formalism, the structure of the interactions is not that simple. We will find a cubic and two quartic vertices (higher vertices are irrelevant at this order) with derivatives acting in various ways. Thus there will be many contributions to the two-loop vacuum diagrams. 

This is spelled out in Sect. 2, where we first briefly review the K\"ahler formalism. Our basic quantum field will be $\wh\f = \frac{1}{2}A_0 \D_0 \f$ where $\f$ is the K\"ahler potential and $\D_0$ the Laplacian in the background metric $g_0$ of area $A_0$. We will refer to $\wh\f$ simply as the K\"ahler field. It will play an important role that this K\"ahler field obviously has no zero-mode.
Let us insist that we are dealing with arbitrary Riemann surfaces. Since a loop expansion is an expansion around a classical solution $(g_0, A_0, \wh\f_0)$, by the Liouville equations of motion the latter may be chosen as $(g_0, A_0,0)$ with $g_0$ being a constant curvature metric with curvature $R_0=\frac{8\pi (1-h)}{A_0}$. However, the metric parametrized by $(g_0,A,\wh\f)$ has arbitrary curvature, of course.
We explicitly write the non-trivial measure for this K\"ahler field $\wh\f$ as it follows from the standard metric on the space of metrics, and write down a first principles formula for the quantum gravity partition function at fixed area, with the action being the Liouville action. We expand the action and measure up to order $\frac{1}{\k^2}$ which corresponds to the two-loop contributions to $\ln Z[A]$. Of course, the measure vertex itself is already a ``one-loop" effect. The expansion of the Liouville action in terms of the K\"ahler field yields the propagator and various $n$-point vertices. For the present two-loop computation we only need to keep the cubic and quartic vertices. We explicitly spell out the expressions arising from the two-loop vacuum diagrams: the setting sun diagram, the figure-eight diagram and the so-called glasses diagram. The non-trivial measure contributes at the same order through a one-loop diagram involving the measure vertex.

Of course, all the expressions must be replaced by the corresponding regularized ones. 
This is the subject of Sect.~3.
Throughout this paper we use the general spectral cutoff regularization developed in \cite{BF}. This amounts to first replacing each propagator $G(x,y)=\sum_n \frac{1}{\l_n}\p_n(x)\p_n^*(y)$ by  $\wh K(t_i,x,y)=\sum_n \frac{e^{-t_i\l_n}}{\l_n}\p_n(x)\p_n^*(y)$ where the $\p_n$ and $\l_n$ are the eigenfunctions and eigenvalues of the relevant differential operator, then substituting $t_i=\frac{\a_i}{\L^2}$ and integrating $\int_0^\infty  \d\a_i \vf(\a_i)$.  Thus, $\L$ is the UV cutoff and $\vf(\a)$ is  a fairly general regulator function. This yields a regularized propagator $G_{\rm reg}(x,y)=\sum_n \frac{f(\l_n/\L^2)}{\l_n}\p_n(x)\p_n^*(y)$
with an almost arbitrary $f$ that is a Laplace transform of the almost arbitrary $\vf$. For large $\L$, the $t_i$ are small and to evaluate the diverging, as well as the finite parts of any diagram, it is enough to know the small $t$ asymptotics of the $\wh K$. Of course, $\wh K(t,x,y)$ is related to the heat kernel $K(t',x,y)$ on the manifold, which has a well-known small $t'$ asymptotic expansion. More precisely $\wh K(t)=\int_t^\infty \d t' K(t')=G-\int_0^t\d t' K(t')$. While we cannot use the small $t'$ asymptotics of $K$ in the first relation, we can use it in the second, which, however, also requires knowledge of the Green's function on the manifold. In most instances, we can satisfy ourselves with the short distance expansion of the latter which, again, is given in terms of the heat kernel coefficients, and the so-called Green's function at coinciding points $G_\zeta(x)$. It is this $G_\zeta$ that captures most of the non-trivial global information about the manifold. In section 3, we briefly recall these results and adapt them to the present two-dimensional case with the zero-modes excluded from the sums. This subtraction of the zero-mode is crucial, of course, but increases even more the number of terms involved in our two-loop computation of $\ln Z[A]$. We end this section by explicitly writing the regularized expressions entering $\ln Z[A]$ and discuss that they can depend on the dimensionful quantities $A$ and $\L$ (as well as the scale $\m$ introduced by the definition of $G_\zeta$) only through the dimensionless combination $A\L^2$ (with no $\m$-dependence).
Incidentally, this argument also explains why it is possible to obtain an explicit form for $\g_{\rm str}$. Indeed, $\g_{\rm str}$ is the coefficient of $\ln A\L^2$ and, hence, it is related to  short-distance singularities which, in turn, are determined by local quantities like heat kernel coefficients.

The regularized loop-integrals then are evaluated in Sect. 4. The general spectral cutoff procedure has the advantage of giving a clear understanding of the divergences that are present. Let us discuss which divergences one might expect in $\ln Z[A]$. First, since the two-loop diagrams only depend on the dimensionless combination $\L^2 A$, there cannot be any area-independent divergence and, a priori, $Z[A]$ must be of the form
\be\label{lnZAform}
\ln Z[A]\big\vert_{\rm 2-loop}=c_1+ c_2 \ln A \L^2 + c_3 A\L^2 +c_4 (\ln A\L^2)^2+ c_5 A \L^2 \ln A\L^2 +{\cal O}(1/\L^2) \ .
\ee
Indeed, inspection of the different contributions shows that no other divergences can occur. The finite coefficients $c_i$ can depend on the regulator functions $\vf(\a)$, as well as on the (global) properties of the Riemann surface.
While all these terms do occur in the individual two-loop diagrams, the (non-local)  $(\ln A\L^2)^2$-terms cancel in $\ln Z[A] : \ c_4=0$. Furthermore, we can always add a cosmological constant counterterm to the action, thus absorbing the $c_3 A \L^2$. Finally, $c_1$ is eliminated by computing instead $\ln \frac{Z[A]}{Z[A_0]}$. This also changes the divergent term $c_2 \ln A \L^2$ into the finite $c_2 \ln\frac{A}{A_0}$. Were it not for the remaining divergent term $\sim c_5$, one could read the two-loop value of $\g_{\rm str}$ from $c_2$. It is interesting to remark that in the sharp-cutoff computation of \cite{BFunpub} the leading divergence was an uncancelled term $\sim N^2 \ln N^2\sim A\L^2 \ln A\L^2$, confirming the presence of a  non-vanishing $c_5 A\L^2 \ln A\L^2$.

This unwanted term $c_5A \L^2 \ln A\L^2 $ is not only divergent, it is also non-local.
Of course, the appearance of a non-local divergence should not be a surprise. In a local quantum field theory, one-loop divergences always are local, but starting at two loops  the divergences are not necessarily local, in particular also due to  the so-called overlapping divergences. However, we know from the standard BPHZ proofs \cite{BPHZ} that they can always be cancelled by one-loop diagrams including local counterterm vertices (as well as tree diagrams including further local counterterm vertices). These counterterm vertices occurring in the one-loop diagrams are themselves determined from the cancellation of the  divergences of the corresponding one-loop $n$-point functions. The same does happen here. This will be worked out in Sect. 5. 

Section 5 is devoted to computing the one-loop two-point function of the K\"ahler field and determining the necessary counterterms to make it finite and regulator independent. This computation is done on an arbitrary Riemann surface and is done consistently in the same spectral cutoff regularization as the computation of the partition function. Note that this two-point function is closely related to the two-point function of the conformal factor $< e^{2\s(x)}e^{2\s(y)}>$. Most of the lengthy computational details are relegated into the appendix. Of course, absence of divergences does not fix the counterterms uniquely. To fix the finite parts of the counterterms requires to impose finite renormalization conditions. While in a massive Minkowski space quantum field theory it is usually convenient to impose conditions on the mass shell, already in a massless theory it is often more convenient to impose the conditions at an arbitrary scale $\m$. In the present theory on a curved Riemann surface, there seems just to be no \emph {natural} finite condition one should impose, rather than any other. Thus there appears to be a whole family of counterterms, depending on two finite parameters. 
Remarkably, for {\it all} choices of these finite parameters, the unwanted $A \L^2 \ln A \L^2$ terms cancel in the two-loop contribution to $\ln Z[A]$, and the coefficient $c_2$ of $\ln A\L^2$ becomes independent of the regulator functions $\vf(\a)\,$! We then get
\begin{multline}\label{ZAresintro}
\ln \frac{Z[A]}{Z[A_0]}\Bigg\vert_{\rm 2-loops\ + \ ct}
=-\frac{1}{\k^2} 
\Big[\wh c_m + 1 +4(1-h) \wh c_R\Big] \ln \frac{A}{A_0} 
+ (A-A_0)\L^2 \left[ \ldots\right] +{\cal O}(1/\L^2) \ .
\end{multline}
Thus, we do indeed find the relation \eqref{KPZscaling}, which is quite remarkable, albeit with an order $\frac{1}{\k^2}$ contribution to $\g_{\rm str}$ that depends on two finite parameters $\wh c_m$ and $\wh c_R$. There exists {\it a choice} for these parameters  which  reproduces the DDK result $\frac{2(1-h)}{\k^2}$, cf. \eqref{gammastrexp}, but it is only \emph{one} choice among others. We will give an argument based on locality favoring the choice $\wh c_m=-1$, which is also the DDK value,  but $\wh c_R$ remains undetermined.
This means that we  actually have an (at least) one-parameter family of quantum gravity theories that we can consistently define in this K\"ahler approach (at least up to the two-loop order of perturbation theory we were studying).

It could well be that some of them become inconsistent when $c>1$, while others are perfectly well-behaved when the $c=1$ ``barrier" is crossed. Since this barrier is invisible at any finite order in the loop-expansion, it is clearly premature to draw any conclusion. It is interesting though, that several equivalent quantization schemes seem to be allowed. In Sect.~6, we end with some remarks about background independence and discuss in some detail the structure that is expected at three loops \cite{LLAB} and beyond.

\subsection{The DDK argument for the scaling of the fixed-area partition function}

Since we will compare our final result with the KPZ relation \eqref{gammastr}, we found it useful to briefly present the DDK argument \cite{DDK} that gives this scaling of the partition function with the area.
This argument is amazingly simple and yields a result that has been cross-checked by other methods, at least for some specific models at certain values of the central charge. On the other hand, as already pointed out, it relies on various simplifying assumptions that cannot be the full truth. 

Instead of using the correct non-trivial measure ${\cal D}\s$, DDK  use a flat free-field measure ${\cal D}_0\s$. At the same time they argue that, in the quantum theory, the coefficient $\k^2$ should be renormalized to $\wt \k^2$ and the definition of the area  can no longer simply be  $\int \d^2 x \sqrt{g_0}\, e^{2\s}$, but that the coefficient in the exponential must also be renormalized so that it becomes $\int \d^2 x \sqrt{g_0}\, e^{2\a\s}$ which should be a conformal primary of weight $(1,1)$ with respect to the Liouville action. Thus
\be\label{ZADDKdef}
Z_{\rm DDK}[A]=\int {\cal D}_0\s \exp\left( -\frac{\wt\k^2}{8\pi} S_L-\m_c^2 A\right) \, 
\dd \Big(A-\int \d^2 x \sqrt{g_0}\, e^{2\a\s}\Big)
 \ .
 \ee

To determine the coefficient $\a$, one can switch to standard normalizations by setting $\wh\s=\wt\k \s$. Then $\frac{\wt\k^2}{8\pi} S_L=\frac{1}{4\pi} \int \d^2 x \sqrt{g_0} \, \big( \frac{1}{2}\wh\s \D_0\wh\s + \frac{\wt\k}{2}R_0 \wh\s\big)$ which represents a standard free-field action with a background charge $\frac{\wt\k}{2}$. The left and right conformal weights of $:e^{2\a\s}:= :e^{\frac{2\a}{\wt\k}\wh\s}:$ then are well-known, and requiring them to equal unity yields
$
-\frac{1}{2}\left(\frac{2\a}{\wt\k}\right)^2 + \frac{\wt\k}{2}\, \frac{2\a}{\wt\k} =1 \ ,
$
with solution
$\a=\frac{\wt\k^2}{4}\left(1-\sqrt{1-\frac{8}{\wt\k^2}}\right)$, where the sign has been chosen to match with the semi-classical limit $\wt\k^2\to\infty$.
The central charge of this ``free" Liouville theory with background charge is
$c_{\rm Liouville}=1+ 3 \wt\k^2$. Background independence requires the total central charge of the matter ($c$), ghosts ($-26$) and the Liouville theory to vanish. This determines
$\wt\k^2=\frac{25-c}{3}=\k^2 -\frac{1}{3}$.

To obtain the area dependence of $Z_{\rm DDK}[A]$ one simply changes the integration variable in \eqref{ZADDKdef} from $\s$ to $\s'=\s-b$ with some constant $b$. The flat measure ${\cal D}_0\s$ is, of course, translationally invariant, while the Liouville action changes as\break\hfill
$S_L[\s]=S_L[\s'] +8\pi b (1-h)$
and the delta scales as
$\dd \Big(A-\int \d^2 x \sqrt{g_0}\, e^{2\a\s}\Big)$\break\hfill
$=e^{-2\a b}\, \dd \Big(e^{-2\a b}A-\int \d^2 x \sqrt{g_0}\, e^{2\a\s'}\Big)$.
Putting things together, and letting  $e^{2\a b}=\frac{A}{A_0}$
one gets
\be\label{ZDDKscale2}
Z_{\rm DDK}[A]=\left(\frac{A}{A_0}\right)^{-1-\frac{\wt\k^2}{2\a}(1-h)}  e^{-\m_c^2 (A-A_0)}Z_{\rm DDK}[A_0]
\ , 
\ee
from which we read
\be\label{gammastr2}
\g_{\rm str}=2-\frac{\wt\k^2}{2\a}(1-h)
=2 + 2(h-1) \frac{ \sqrt{25-c}}{\sqrt{25-c}-\sqrt{1-c}} \ .
\ee


\section{The K\"ahler approach}

Consider a compact Riemann surface with fixed complex structure moduli. Up to diffeomorphisms, any metric $g$ on the surface can be written in conformal gauge as $g=e^{2\s} g_0$ where $g_0$ is a reference metric that we choose as the constant curvature metric  associated with some area $A_0$.
Rather than writing the two-dimensional metric in terms of this $g_0$ and the conformal factor $\s$ one can also parametrize it in terms of $g_0$, the area $A$ and the K\"ahler potential $\f$ as follows
\be\label{gg0sig}
g = e^{2\sigma}g_{0} \quad , \quad
e^{2\s}= \frac{A}{A_0}\left(1-\frac{1}{2} A_0 \D_0\f\right)\ ,
\ee
where $\D_0$ denotes the Laplacian for the metric $g_0$. In general, $\D$ will always denote the {\it positive} Laplacian, i.e. $\D=-g^{ij} \nabla_i \nabla_j$. We have, of course, $\D=e^{-2\s} \D_0$. Clearly, $A$ is the area associated with $g$.
This K\"ahler paramatrization \eqref{gg0sig} has certain advantages and is certainly most convenient if one wants to consider metrics of fixed area. 
Given $\sigma$, the above relation actually defines $A$ and $\phi$ uniquely, up to unphysical constant shifts of $\phi$. 
The relation \eqref{gg0sig} is equivalent to the relation
\be\label{kahlform} \omega =\frac{A}{A_{0}} \omega_{0} + i A\partial\bar\partial\phi\ee
between the volume (K\"ahler) forms $\o$ and $\o_0$ of the metrics $g$ and $g_{0}$. Often we will use a rescaled (constant curvature)  metric $g_*$ of area $A$, with corresponding Laplace operator $\D_*$ and Ricci scalar $R_*$ given by
\be\label{gstar}
g_*=\frac{A}{A_0} g_0 \quad  ,\quad  \D_*=\frac{A_0}{A}\D_0 
\quad  ,\quad  R_*=\frac{A_0}{A} R_0=\frac{8\pi (1-h)}{A} \ .
\ee 
In particular, since $A_0\D_0 = A \D_*$, eq.~\eqref{gg0sig} can also be written as
\be\label{gg0sig2}
e^{2\s}= \frac{A}{A_0}\left(1-\frac{1}{2} A \D_*\f\right)
\quad \Leftrightarrow \quad g=g_* \left(1-\frac{1}{2} A \D_*\f\right)\ .
\ee
Obviously, one has
\be\label{obvious}
\int\d^2 x\, \sqrt{g} =\int \d^2 x\, \sqrt{g_*}=A \quad , \quad
\int \d^2 x\, \sqrt{g_*}\, R_*= 8\pi (1-h) \ .
\ee
As compared to the DDK approach, where in the quantum theory the area is computed as $\int\d^2 x \sqrt{g_0} e^{2\a\s}$ (cf \eqref{ZADDKdef}), in the present K\"ahler approach, the area $A$ is a ``coordinate" on the space of metrics and \eqref{obvious} always holds exactly even in the quantum theory where $\f$ or rather $A\D_* \f$ is the quantum field.

Throughout this present paper, we will only use the Liouville action that is expressed in terms of $\s$ and not the Mabuchi action (as was done in \cite{BFK}) that involves $\f$ directly. For this reason, we define what we call the K\"ahler field $\wh\f$ as
\be\label{phihatdef}
\wh\f=\frac{1}{2} A \D_* \f \ .
\ee
Since $\f$ is defined only up to constant shifts, the relation between $\f$ and $\wh\f$ is one-to-one. Eq.~\eqref{gg0sig2} now simply reads
\be\label{gg0sig3}
e^{2\s}= \frac{A}{A_0}\left(1-\wh\f\right)
\quad \Leftrightarrow \quad g=g_* \left(1-\wh\f\right)\ .
\ee
Obviously, positivity of the metric implies  the non-perturbative constraint $\wh\f<1$. 

\subsection{The measure on the space of metrics}\label{measureSec}

In quantum gravity we will need to integrate over the space of metrics modulo diffeomorphisms. The integration measure on this space can be derived from a choice of metric on the space of metrics. It is generally assumed that this metric should be ultralocal and, hence, of the form $||\dd g||^2=\int\d^2 x \sqrt{g}\ \dd g_{ab}\dd g_{cd}\, \big( g^{ac} g^{bd}+c\, g^{ab}g^{cd}\big)$ for some constant $c>-1/2$. Using $g=e^{2\s}g_0$, this yields $||\dd g||^2=8 (1+2c)  \ ||\dd\s||^2$ with
\be\label{sigmametric}
||\dd\s||^2=
\int\d^2 x \sqrt{g_0}\ e^{2\s}\,(\dd\s)^2 \ .
\ee 
The integration measure ${\cal D}\s$ over $\s$ is determined from this metric.  It is {\it not} the measure of a free field, because of the non-trivial factor $e^{2\sigma}$. Instead of $\sigma$, we will use equivalently the variables $(A,\phi)$. Inserting \eqref{gg0sig} into \eqref{sigmametric} yields
\be\label{ddsigmaddphi}
||\dd\s||^2=\frac{(\dd A)^2}{ 4 A} + ||\dd\s||^2_A \ ,
\ee
with the $||\dd\s||^2_A$ being the metric on the space of metrics for fixed area $A$, see \cite{BFK},
\be\label{sigmametric2}
||\dd\s||^2_A =\frac{1}{16} \int\d^2 x \sqrt{g}\, (A\D \dd\f)^2
=\frac{1}{4} \int\d^2 x \sqrt{g_*}\, (1-\wh\f\,)^{-1}\, {\dd\wh\f\,}^2
\, .\ee
%
%
%
Formally, \eqref{ddsigmaddphi} and \eqref{sigmametric2} thus induce a measure
\be\label{measure3}
\mathcal D\sigma 
=\frac{\d A}{ \sqrt{A}}\, \Bigl[\Det' \bigl( 1-\wh\f \,\bigr)^{-1}\Bigr]^{1/2} \, {\cal D}_*\wh\f \ .
\ee
%
%
where ${\cal D}_*\wh\f$ is the standard free field integration measure in the background metric $g_{*}$ deduced from the metric $||\dd\wh\f||_{*}^2=\int\!\d^2 x \sqrt{g_*}\, \dd\wh\f^2$. The notation $\Det'$ means that we are not taking into account the zero mode  when computing the determinant, consistently with the fact that  $\wh\phi$ has no zero-mode.
The measure ${\cal D}_*\wh\f$ can be expressed in the traditional way by expanding $\wh\f$ in eigenmodes of the Laplace operator $\D_*$. We denote by $0=d_{0}<d_{1}\leq d_{2}\leq\cdots$  the eigenvalues of $\D_*$ and by $\p_r$ the eigenfunctions which we choose to be real. Then
\be\label{eigenexp}
\wh\f=\sum_{r>0} c_r \p_r\, , \quad
\D_* \p_r=d_r\p_r\, , \quad
\int\!\d^2 x \sqrt{g_*} \,\p_r \p_s= \delta_{rs}\, ,
\ee
and the measure is defined as
\be\label{measureA}
{\cal D}_* \wh\f=\prod_{r>0} \d c_r \ .
\ee
%
%
%
%
%
Our starting point for the computation of the quantum gravity partition function at fixed area then is (cf \eqref{ZAdef})
\ba\label{ZAKahlerdef}
Z[A]&=&\frac{e^{-\m_c^2 A}}{\sqrt{A}}\, \int  {\cal D}_*\wh\f\, 
\Bigl[\Det' \bigl( 1-\wh\f \,\bigr)^{-1}\Bigr]^{1/2} \,
\exp\left(-\frac{\k^2}{8\pi} S_L\Big[\s[A,\wh\f]\Big] \right) \nonumber\\
&\equiv&  \frac{e^{-\m_c^2 A}}{\sqrt{A}}\, \int  {\cal D}_*\wh\f\, 
\exp\left(-S_{\rm measure}-\frac{\k^2}{8\pi} S_L\Big[\s[A,\wh\f]\Big] \right)  \ .
\ea

A few remarks are in order: first, writing the metric $g$ in the form \eqref{gg0sig} amounts to fixing the action of the diffeomorphisms. This produces the well-known ghosts, whose effect is to produce the $26$ in the coefficient $\kappa^{2}$ of the Liouville action. A further subtlety arises in the case of the sphere, $h=0$, because the gauge-fixing \eqref{gg0sig} then is incomplete, as discussed in \cite {BFK}. An additional gauge fixing of the residual $\text{SL}(2,\mathbb C)/\text{SU}(2)$ group of diffeomorphisms acting non-trivially on $\sigma$ and $\wh\phi$ must be performed. The result  is  to project out the spin-one modes of $\wh\phi$ in the decomposition \eqref{eigenexp} and to produce an overall factor of $A^{3/2}$ in the partition function coming from the Faddeev-Popov determinant. In the rest of this paper, we will implicitly assume that $h>0$, in order not to explicitly deal with this complication.

Second, as already mentioned, positivity of the metric constrains the space of K\"ahler fields $\wh\f$ over which we integrate with the measure \eqref{measure3} to the subspace $\wh\f<1$. This constraint is irrelevant in perturbation theory and thus will not bother us in the present paper. 

Third, using $\f$ instead of $\wh\f$ as the basic integration variable  generates a Jacobian determinant that cancels a similar determinant at one loop coming from the Liouville action \cite{BFK}. This cancellation could be incomplete if a so-called multiplicative anomaly is present. The correct cancellation thus depends on the regularization scheme. As was discussed in  detail in \cite{BFK},  when using the spectral regularization  no multiplicative anomaly occurs and there is no subtlety associated with the change of integration variables from $\f$ to $\wh\f$.

\subsection{Two-loop expansion of the Liouville action and measure}

As is well-known, the classical saddle points of the Liouville action \eqref{Liouvaction}  are the constant curvature metrics of arbitrary area $A$. Thus it is particularly convenient to take the background metric $g_0$ to be a constant curvature metric of given area $A_0$ as anticipated above. The classical solutions $\s_{\rm cl}$ then simply are the constants $e^{2\s_{\rm cl}}=\frac{A}{A_0}$.
The Liouville action may then be trivially rewritten in terms of $\s$ and the rescaled $g_*$, $\D_*$, $R_*$ as
\be\label{Liouv*action}
S_L[\s]=\int\d^2 x\, \sqrt{g_*} \big( \s \D_* \s+R_* \s\big) \  ,
\ee
and 
\eqref{gg0sig2} becomes
\be\label{sigsigclass}
\s-\s_{\rm cl} = \frac{1}{2} \ln(1-\wh\f) \quad , \quad  \s_{\rm cl}=\frac{1}{2} \ln\frac{A}{A_0} \ .
\ee
Since the first term in $S_L$ is not affected by constant shifts of $\s$ and the second term is linear, one obviously has
\be\label{Liouv2}
S_L[\s]=S_L[\s_{\rm cl}] + S_L [\s-\s_{\rm cl}]
=4\pi (1-h) \ln\frac{A}{A_0} + S_L \big[{\textstyle \frac{1}{2}} \ln(1-\wh\f)\big] \ .
\ee
Expanding the logarithm we get
\begin{multline}\label{Liouvexp1}
S_L[\s-\s_{\rm cl}]=\int \d^2 x\, \sqrt{g_*}\Big[
\frac{1}{4}\wh\f (\D_*-R_*)\wh\f +\frac{1}{4} {\wh\f\,}^2(\D_*-\frac{2}{3}R_*)\wh\f \\
+\frac{1}{16}{\wh\f\,}^2 \D_*{\wh\f\,}^2 +\frac{1}{6}{\wh\f\,}^3\D_*\wh\f -\frac{1}{8} R_* {\wh\f\,}^4
+{\cal O}({\wh\f\,}^5) \Big] \ .
\end{multline}
As is clear from the definition of the quantum gravity partition function \eqref{ZAKahlerdef}, the quantity $\frac{1}{\k^2}$ is a loop-counting parameter, i.e. the loop-expansion of $Z[A]$ is an expansion in powers of $\frac{1}{\k^2}$. To see this clearly, we rescale $\wh\f$ as
\be\label{phitildedef}
\wt\f=\frac{\k}{\sqrt{16\pi}} \, \wh\f \ ,
\ee
so that
\begin{multline}\label{Liouvexp2}
\frac{\k^2}{8\pi} S_L[\s]=\frac{\k^2}{2}(1-h)\ln\frac{A}{A_0} 
+\int \d^2 x\, \sqrt{g_*} \, \frac{1}{2} \wt\f (\D_*-R_*) \wt\f \\
+\int \d^2 x\, \sqrt{g_*} \Big[
\frac{\sqrt{4\pi}}{\k} {\wt\f}^2(\D_*-\frac{2}{3}R_*)\wt\f 
+\frac{2\pi}{\k^2}\wt\f^2 \D_*\wt\f^2 \\
+\frac{16\pi}{3\k^2}\wt\f^3\D_*\wt\f -\frac{4\pi}{\k^2} R_* \wt\f^4
+{\cal O}(\k^{-3}) \Big] \ .\hskip1.7cm
\end{multline}
The first term in the first line provides the classical contribution to the partition function. The second term of the first line yields the one-loop determinant studied in \cite{BFK}. It also provides a standard propagator for the present two-loop computation. The terms of the second and third line provide the cubic and quartic vertices relevant for the two-loop vacuum diagrams. Clearly, the ${\cal O}(\k^{-3})$ terms correspond to quintic and higher vertices that can only contribute to three (and higher)-loop vacuum diagrams.

The non-trivial measure does not contribute at one-loop. However, it gives a two-loop (as well as higher loop) contribution. From \eqref{ZAKahlerdef}
we get
\be\label{Smeasure}
S_{\rm measure}=-\frac{1}{2} \ln \Det'(1-\wh\f)^{-1}=\frac{1}{2}{\rm Tr}'\ln(1-\wh\f)
={\rm Tr}' \left(-\frac{\sqrt{4\pi}}{\k} \wt\f -\frac{4\pi}{\k^2}\wt\f^2 +{\cal O}(\k^{-3})\right) \ .
\ee
More explicitly, in terms of the orthonormal set of eigenfunctions $\p_r$ of the Laplace operator, cf \eqref{eigenexp}, the trace of any operator $O$ is given by
\be\label{tracedef}
{\rm Tr}' O=\sum_{r>0} \langle\p_r| O |\p_r\rangle
=\int\d^2 x \sqrt{g_*}\sum_{r>0} \langle\p_r|x\rangle \langle x| O |\p_r\rangle
=\int\d^2 x \sqrt{g_*}\sum_{r>0} \p_r^2(x)\, O(x) \ ,
\ee
(we have chosen real eigenfunctions $\p_r$) and, hence,
\be\label{Smeasure2}
S_{\rm measure}=\int\d^2 x \sqrt{g_*}\sum_{r>0} \p_r^2(x) \left(-\frac{\sqrt{4\pi}}{\k} \wt\f(x) 
-\frac{4\pi}{\k^2}\wt\f^2(x) +{\cal O}(\k^{-3})\right) \ .
\ee
Of course, $\sum_{r>0} \p_r^2(x)=\frac{\dd(0)}{\sqrt{g_*(x)}}-\p_0^2$ is divergent and needs to be regularized. As we will show below, after regularization, $\sum_{r>0} \p_r^2(x)$ will not depend on $x$ and, since $\wt\f$ has no zero-mode, the term linear in $\wt\f$ will drop out (cf \eqref{Ktildexx} and \eqref{nolinearterm} below). Thus $S_{\rm measure}$ provides a  quadratic vertex with a diverging coefficient. This is very similar to a counterterm. We will have more to say about this analogy later-on.

The reader might wonder what is the advantage of rewriting the simple-looking Liouville action \eqref{Liouv*action} in terms of $\wh\f$ or $\wt\f$ which has resulted in a complicated action \eqref{Liouvexp1} or \eqref{Liouvexp2} with cubic and quartic (and higher)  interactions that all involve derivatives. Would it not be simpler to use 
$\frac{1}{2} \ln(1-\wh\f) =\s-\s_{\rm cl}$ as the basic field instead? One important point concerns the zero-mode. The absence of zero-mode is easy to implement for $\wh\f$, while for $\s-\s_{\rm cl} =-\frac{1}{2} \wh\f-\frac{1}{4}\wh\f^2 -\frac{1}{8}\wh\f^3-\ldots$ it results in a very complicated constraint. Taking this constraint properly into account is highly non-trivial and probably equivalent in difficulty to working with the complicated actions  \eqref{Liouvexp1} or \eqref{Liouvexp2}.

\vskip1.cm
\subsection{The two-loop contributions to the quantum gravity partition function}

Of course, it is not the partition function $Z[A]$  itself but its logarithm $W[A]=\ln Z[A]$  -- which is the sum of connected vacuum diagrams -- that has a convenient loop expansion:
\be\label{WLexp}
\ln Z[A]\equiv W[A]=\sum_{L\ge 0} W^{(L)}[A] \,
\ee
with 
\be\label{W0}
W^{(0)}[A]=\frac{\k^2}{2}(h-1)\ln\frac{A}{A_0} \,
\ee
as can be read off directly from \eqref{Liouvexp2}. The one-loop contribution was computed in \cite{BFK} using the spectral cutoff regularization. There, a more general action, including also the Mabuchi action was considered, so one had to use $\f$ rather than $\wh\f$ as a basic integration variable. The result was given in terms of various regularized determinants. With the Liouville action only, the result  simply becomes
\be\label{W1}
W^{(1)}[A]=-\frac{1}{2}\ln\frac{A}{A_0} -\frac{1}{2} \ln \Det'(\D_*-R_*)+\wt c_1
=\frac{1-7h}{6} \ln\frac{A}{A_0}  + \mu^2_{\rm c,div} A +c_1\ ,
\ee
where the divergent piece $\mu^2_{\rm c,div} A\sim \L^2 A$ is cancelled by the renormalization of the cosmological constant $\mu^2_{\rm c}$, and $c_1$ and $\wt c_1$ are $A$-independent finite constants that could be eliminated by computing $W^{(1)}[A]-W^{(1)}[A_0]$  instead. The coefficients of $\ln\frac{A}{A_0}$ yield the contributions to $\g_{\rm str}-3$ and thus from \eqref{W0} and \eqref{W1} 
\be\label{gstrK01}
\g_{\rm str}=\frac{h-1}{2} \k^2 +\frac{19-7h}{6} +{\cal O}(\k^{-2}) \ ,
\ee
in agreement with \eqref{gammastrexp}. As already mentioned, one of the present goals is to determine the order $\frac{1}{\k^2}$ term in this expansion which comes from the two-loop contribution.

The two-loop contribution to $W[A]$  is
\be\label{W2}
W^{(2)}[A]=\sum \Big[\text{connected vacuum diagrams  $\sim \frac{1}{\k^2}$}\Big] \ .
\ee
This includes the genuine two-loop diagrams made with the vertices of $S_L$, as well as a one-loop diagram made with the vertex from $S_{\rm measure}$ and, as we will see, also one-loop diagrams made with further counterterm vertices. 

\subsubsection{Vertices}

The Feynman rules are straightforwardly read from the expansions \eqref{Liouvexp2} and \eqref{Smeasure2}.
Note that we normalize our vertices without including any symmetry factors (i.e. a term $\a \wt \f^n$ in the total action gives a vertex with $n$ legs and a factor $- \a$, {\it not} $-\a n!$) so that when evaluating the diagrams one has to count all possible contractions.
There is a cubic vertex 
\def\diagramcubic{\begin{picture}(26,10)(0,0)
\put(10,3.5){\line(1,1){10}}\put(10,3.5){\line(-1,1){10}}
\put(10.5,3.5){\line(0,-1){10}}\put(9.5,3.5){\line(0,-1){10}}\end{picture}}
\def\diagramquarticone{\begin{picture}(26,10)(0,0)
\put(10,3.5){\line(-1,1){10}}\put(10,3.5){\line(-1,-1){10}}
\put(10,4.){\line(1,0){5}}\put(10,3.0){\line(1,0){5}}
\put(15,3.5){\line(1,1){10}}\put(15,3.5){\line(1,-1){10}}\end{picture}}
\def\diagramquartic2{\begin{picture}(26,10)(0,0)
\put(10,3.5){\line(1,1){10}}\put(10,3.5){\line(-1,1){10}}
\put(10,3.5){\line(-1,-1){10}}\put(9.5,3.5){\line(1,-1){10}}\put(10.5,3.7){\line(1,-1){10}}\end{picture}}
\def\diagrammeasure{\begin{picture}(26,10)(0,0)
\put(10,3.5){\line(1,-1){10}}\put(10,3.5){\line(1,1){10}}
\put(10,3.5){\circle{2}}\put(10,3.5){\circle{1}}\put(10,3.5){\circle{3}}\end{picture}}
\def\diagrammeasureloop{\begin{picture}(26,10)(0,0)
\put(10,3.5){\circle{2}}\put(10,3.5){\circle{1}}\put(15,3.5){\circle{10}}\end{picture}}
\def\diagrammeasureloopsmall{\begin{picture}(10,10)(0,0)
\put(2.5,3.5){\circle{2}}\put(2.5,3.5){\circle{1}}\put(5,3.5){\circle{5}}\end{picture}}
\be\label{cubic}
\diagramcubic = \ \ -\frac{\sqrt{4\pi}}{\k}(\D_*-\frac{2}{3}R_*) \ ,
\ee
where $(\D_*-\frac{2}{3}R_*) $ is meant to act on the propagator connected to the fat line of the vertex. There are also two quartic vertices,
\be\label{quartic1}
\diagramquarticone\qquad = \ \ -\frac{2\pi}{\k^2}(\D_*-2 R_*) \ ,
\ee
with $(\D_*-2 R_*)$ acting on the {\it two} propagators connected to the fat part of the vertex (of course, upon integrating by parts, it does not matter whether one chooses the two lines to the left or the two lines to the right),
and
\be\label{quartic2}
\diagramquartic2\ \  = \ \ -\frac{16\pi}{3\k^2}\D_*\ .
\ee
The measure yields
\be\label{measurevertex}
\diagrammeasure\ \ =\ \  \frac{4\pi}{\k^2} \sum_{r>0}\p_r^2(x) \ .
\ee
As already mentioned, the linear term in $S_{\rm measure}$ drops out,  after regularization, as we will discuss below, see \eqref{nolinearterm}.

\subsubsection{Propagator}

These vertices are connected by propagators that are 
\be\label{Greenfct}
\wt G(x,y)=\langle x |(\D_*-R_*)^{-1}|y\rangle' \ .
\ee
The tilde on $G$ and the prime on the r.h.s. indicate that the zero-mode is not to be included. This propagator can be written explicitly in terms of the eigenvalues $\l_r$ and eigenfunctions $\p_r$ of 
\be\label{Dop}
D=\D_*-R_* \quad , \quad D \p_r = \l_r \p_r \ .
\ee
Since  $R_*=\frac{8\pi(1-h)}{A}$ is constant,  $D$ and $\D_*$ have the same (real) eigenfunctions $\p_r$, while the eigenvalues simply are related by $\l_r=d_r-R_*$.
Since $\wt\f$ has no zero-mode, the propagator is given by the sum over all non-zero modes as
\be\label{Green2}
\wt G(x,y)=\sum_{r>0} \frac{\p_r(x)\p_r(y)}{\l_r} \ .
\ee
Note that the zero-mode of the Laplace operator $\D_*$ always is a constant and from the normalization we then get 
\be\label{zero-mode}
\p_0=\frac{1}{\sqrt{A}} \ .
\ee
Furthermore, for genus $h>1$ one has $\l_0=-R_* >0$, and then one can explicitly subtract the zero-mode contribution $\frac{\p_0^2}{\l_0}=\frac{1}{8\pi(h-1)}$\ :
\be\label{Green3}
\text{for $h>1$ :}\quad
\wt G(x,y)=G(x,y) -\frac{1}{8\pi(h-1)}
\quad , \quad G(x,y)=\sum_{r\ge0} \frac{\p_r(x)\p_r(y)}{\l_r}.
\ee
Note also that for all $h\ge 1$, we have $\l_r>0$ for $r>0$ since the eigenvalues $d_r$ of the Laplacian are positive (for $r>0$) and $-R_*\ge 0$. It is only for $h=0$ where $\l_r\equiv \l_{l,m}=\frac{4\pi}{A}l(l+1)-\frac{8\pi}{A}$ that we get $\l_{0,0}<0$ and $\l_{1,m}=0$. As already mentioned, for $h=0$, these three spin-1 modes are excluded by the ${\rm SL}(2,{\bf C})/{\rm SL}(2,{\bf R})$ gauge fixing. In the sequel we will always implicitly assume $h\ge 1$.
While $G(x,y)$ satisfies $D_x G(x,y) = \frac{1}{\sqrt{g_*}}\dd(x-y)$, $\wt G(x,y)$ satisfies
\be\label{Gtilderel}
D_x \wt G(x,y)=\frac{1}{\sqrt{g_*}}\dd(x-y) - \frac{1}{A} \ .
\ee
Obviously $\wt G(x,y)$ and $G(x,y)$ are symmetric under exchange of $x$ and $y$.
We will give explicit expressions for the short-distance asymptotics in the next section.

Of course, when computing the Feynman diagrams, every coordinate $x_i$ associated with a vertex is to be integrated with $\int \d x_i \equiv\int \d^2 x_i \sqrt{g_*(x_i)}$.

\subsubsection{The vacuum diagrams of order $1/\k^2$}

The two quartic vertices both give a ``figure-eight" diagram $\diagramcbig$ with the four lines of a single vertex connected by two propagators. The cubic vertex gives rise to a ``setting sun" diagram $\diagrambbig$ with two cubic vertices connected by three propagators. The cubic vertex also give a ``glasses" diagram $\diagramdbig$ with the two vertices joined by a single propagator and the remaining two lines of each vertex connected by a propagator. Finally the measure vertex gives the ``measure" (one-loop) diagram $\diagrammeasureloop$ with the two lines of the vertex connected by a single propagator.

\noindent
\underline{The setting sun diagram :}
This diagram actually gets two contributions, one with the two fat lines of the two cubic vertices connected by the same propagator (two possible contractions) and one with the two fat lines not connected by the same propagator (four contractions). Overall, there is also a factor $\frac{1}{2}$ from expanding $e^{-S_{\rm int}}$ to second order. Thus
\begin{multline}\label{settingsun1}
\diagrambbig=\frac{4\pi}{\k^2} \int\d^2 x \sqrt{g_*(x)}\d^2 y \sqrt{g_*(y)}\,
\wt G(x,y) \Bigg[ \wt G(x,y) (\D_*^x-\frac{2}{3}R_*)(\D_*^y-\frac{2}{3}R_*)\wt G(x,y) \\
+ 2 \big[(\D_*^x-\frac{2}{3}R_*)\wt G(x,y) \big](\D_*^y-\frac{2}{3}R_*)\wt G(x,y) \Bigg] \ .
\end{multline}

\noindent
\underline{The glasses diagram :}
This diagram gets four contributions, since for each of the cubic vertices the fat lines can either be contracted with a line from the same vertex (giving a factor of 2) or with a line of the other vertex (factor 1). Again, overall, there is also a factor $\frac{1}{2}$. Thus
\begin{multline}\label{glasses1}
\diagramdbig=\frac{2\pi}{\k^2} \int\d^2 x \sqrt{g_*(x)}\d^2 y \sqrt{g_*(y)}\,
\Bigg[  (\D_*^x-\frac{2}{3}R_*)\wt G(x,x) + 2  (\D_*^x-\frac{2}{3}R_*) \wt G(x,z)\big\vert_{z=x}\Bigg]\\
\times \wt G(x,y)
\Bigg[  (\D_*^y-\frac{2}{3}R_*)\wt G(y,y) + 2  (\D_*^y-\frac{2}{3}R_*) \wt G(y,z)\big\vert_{z=y}\Bigg] \ .
\end{multline}

\noindent
\underline{The figure-eight diagram :}
This diagram gets three contributions : one from the quartic vertex \eqref{quartic2} and two from the different ways to contract the lines of the quartic vertex \eqref{quartic1}. Taking into account the different numbers of contractions in each case (3, 2 ond 1) yields\begin{multline}\label{figeight1}
\diagramcbig=-\frac{8\pi}{\k^2} \int\d^2 x \sqrt{g_*(x)}\,
\Bigg[ \frac{1}{4} \wt G(x,x) (\D_*^x-2R_*)\wt G(x,x) \\
+ \frac{1}{2} \Big[  (\D_*^x-2R_*) \big(\wt G(x,z)\big)^2\Big]\Big\vert_{z=x}
+ 2 \wt G(x,x) \big[ \D_*^x \wt G(x,z)\big]\Big\vert_{z=x} \Bigg] \ .
\end{multline}

\noindent
\underline{The measure diagram :}
The vertex \eqref{measurevertex} simply gives
\be\label{measurediag}
\diagrammeasureloop=\frac{4\pi}{\k^2} \int\d^2 x \sqrt{g_*(x)}\, \sum_{r>0}\p_r^2(x) \wt G(x,x) \ .
\ee

In addition to these diagrams there are also one-loop diagrams involving counter\-term vertices that contribute at the same order in $\frac{1}{\k^2}$. We will discuss them in Sect.~5. So far, we can summarize
\ba\label{W2-2}
W^{(2)}[A]&=&W^{(2)}[A]\big\vert_{\rm loops}+W^{(2)}[A]\big\vert_{\rm ct} \\
W^{(2)}[A]\big\vert_{\rm loops}&=&
\diagrambbig \ +\  \diagramdbig\  +\  \diagramcbig\ + \diagrammeasureloop 
\\
W^{(2)}[A]\big\vert_{\rm ct}&=&\  \text{counterterm contributions at order } \frac{1}{\k^2} \ .
\ea

\vskip1.cm
\section{Spectral cutoff regularization}

\subsection{The spectral cutoff regularization scheme}

Of course, the above expressions for the Feynman diagrams are formal and have to be regularized in a specific consistent way.
A convenient and powerful regularization scheme is provided by the general spectral cutoff approach developed in \cite{BF}. At one loop it generalizes the well-known abstract zeta function regularization while providing physical  motivation and intuition. At higher loops it amounts to regulating the propagators in a specific way. A basic feature of the spectral cutoff is to replace any sum over the eigenvalues $\l_r$ of the relevant differential operator (which at present is $D=\D_*-R_*$) by a regulated sum as
\be\label{basicreg}
\sum_r F(\l_r) \to \Big[\sum_r F(\l_r)\Big]_{\vf,\L,M}=\int_0^\infty \d\a\, \vf(\a) \sum_r e^{-\frac{\a}{\L^2} (\l_r+M^2)} F(\l_r) \ .
\ee
The regulator function $\vf$ is relatively arbitrary, except for the normalization condition  $\int_0^\infty \d\a \vf(\a)=1$ (which ensures that for $\L\to\infty$ one recovers the unregulated sum), and certain regularity properties as $\a\to 0$ or $\a\to\infty$. An important physical requirement is that, in the end, all $\vf$ or $M$-dependence should be only in terms that can be changed by the addition of (local) counterterms, while any physical part should be regulator independent. This has been checked on several examples in \cite{BF}. At present, there is no need to introduce the constant $M$ and we set $M=0$.
Then, in particular, the regulated propagator is
\be\label{propreg}
\wt G(x,y) \to \Big[ \wt G(x,y)\Big]_{\vf,\L} =\int_0^\infty \d\a\, \vf(\a) \sum_{r>0} e^{-\frac{\a}{\L^2} \l_r}\  \frac{\p_r(x)\p_r(y)}{\l_r} \ .
\ee
The right-hand-side is actually related to the heat-kernel on the manifold. Indeed, the heat kernel is defined as
\be\label{heatkernel}
K(t,x,y)=\sum_{r\ge0} e^{-\l_r t} \,\p_r(x) \p_r(y) \ .
\ee
Here, we are always interested in the corresponding quantities excluding the zero-modes, which we indicate by adding a tilde:
\be\label{heatkerneltilde}
\wt K(t,x,y)=\sum_{r>0} e^{-\l_r t}\, \p_r(x) \p_r(y) =K(t,x,y)-\frac{e^{R_* t}}{A}\ .
\ee
Integration with respect to $t$ yields the ``hatted heat kernel" without zero-mode:
\be\label{Khattilde}
\wh{\wt K}(t,x,y)=\int_t^\infty \d t'\, \wt K(t',x,y) = \sum_{r>0} \frac{e^{-\l_r t}}{\l_r} \p_r(x) \p_r(y) \ .
\ee
The integration is convergent at $+\infty$ since $\l_r>0$ for all $r>0$.
Of course, $\wt K(t,x,y)$ and $\wh{\wt K}(t,x,y)$ are symmetric under exchange of $x$ and $y$ and one has the following relations
\be\label{KKhatrel}
-\frac{\d}{\d t}\wh{\wt K}(t,x,y)= D_x \wh{\wt K}(t,x,y)= D_y \wh{\wt K}(t,x,y)= {\wt K}(t,x,y) \ ,
\ee
as well as
\be\label{KhatGrel}
\wt G(x,y)=\wh{\wt K}(0,x,y) \ .
\ee
As is clear from the definitions, for $t>0$, $\wt K(t,x,y)$ and $\wh{\wt K}(t,x,y)$ are given by converging sums and are finite, even as $x\to y$. For $t\to 0$ one recovers various divergences, in particular $\wh{\wt K}(t,x,y)$ then yields the short distance singularity of $\wt G(x,y)$ which is well-known to be logarithmic. An important quantity, called the ``Green's function at coinciding points", is obtained by subtracting this short distance singularity and taking $x\to y$. More precisely, we define, cf \cite{BF}
\be\label{Gzetadef}
\wt G_\zeta(y)=\lim_{x\to y} \left[ \wt G(x,y) + \frac{1}{4\pi}\left( \ln\frac{\ell^2_A(x,y) \m^2}{4} +2\g\right)\right]
\ ,
\ee
where $\ell^2_A(x,y)$ is the geodesic distance between $x$ and $y$ in the metric of area $A$, and $\m$ is an arbitrary scale.

Since $\wt K(t,x,y)$ and $\wh{\wt K}(t,x,y)$ do not include the zero-mode, their integrals over $x$ or over $y$ vanish (as is also the case for $\wt G(x,y)$, of course):
\be\label{intxvanishes}
\int\d^2 x \sqrt{g_*(x)} \, \wt K(t,x,y)= \int\d^2 x \sqrt{g_*(x)} \, \wh{\wt K}(t,x,y) =0 \ .
\ee
Also
\be\label{delta-0}
\wt K(0,x,y)=\frac{\delta(x-y)}{[g_*(x)g_*(y)]^{1/4}} -\frac{1}{A} \ .
\ee

The regularized Green's function \eqref{propreg} is now seen to be given by
\be\label{propreg2}
\Big[ \wt G(x,y)\Big]_{\vf,\L} =\int_0^\infty \d\a\, \vf(\a)\, \wh{\wt K}(\frac{\a}{\L^2},x,y) \ .
\ee
If a given Feynman diagram  (integral) $I_n$ contains $n$ propagators, we can now define its regularized version as
\be\label{Feynmanreg}
I_n^{\rm reg} \equiv \Big[ I_n \Big]_{\vf,\L} = \left(\prod_{i=1}^n \int_0^\infty \d\a_i\, \vf(\a_i) \right)\ I_n(t_1=\frac{\a_1}{\L^2},\ldots, t_n=\frac{\a_n}{\L^2}) \ ,
\ee
where $I_n(t_1, \ldots, t_n)$ is the Feynman diagram (integral) with all propagators $\wt G(x_i,y_i)$ replaced by $\wh{\wt K}(t_i,x_i,y_i)$. It is obvious from \eqref{Feynmanreg} that the only  part of $I_n(t_i)$ that contributes is the part that is completely symmetric in all $t_i$.

The strategy is to compute the relevant $I_n(t_1, \ldots, t_n)$ and extract the small $t_i$ asymptotics. Since we always let
\be\label{tialphai}
t_i=\frac{\a_i}{\L^2}
\ee
the small $t_i$ and large $\L$ asymtotics are, of course, equivalent. To do this, a basic tool is the well-known small $t$ asymptotics of the heat kernel $K(t,x,y)$, resp $\wt K(t,x,y)$. Unfortunately, this does not allow us to get the small $t$ asymptotics of $\wh{\wt K}(t,x,y)$ since by \eqref{Khattilde} the latter involves $\wt K(t',x,y)$ for all $t'\ge t$ which are not all small. However, using \eqref{Khattilde} and \eqref{KhatGrel} we can write
\be\label{KhatGint}
\wh{\wt K}(t,x,y)=\wt G(x,y) - \int_0^t \d t'\, \wt K(t',x,y) \ .
\ee
In the second term $t'\le t$ is small for small $t$ and we get a useful formula if we can also say something about the un-regularized Green's function $\wt G(x,y)$. This will be the case if  $x$ is close to $y$, where a short distance asymptotics is available.

Later-on, we will need to compute integrals over products of $\wt K$ and $\wh{\wt K}$. Some of these integrals follow straightforwardly from the definitions \eqref{heatkerneltilde} and \eqref{Khattilde} and the orthonormality of the eigenfunctions $\p_n$ :
\be\label{kkhatint}
\int\d^2 x \sqrt{g(x)}\, \wt K(t_1,u,x)\, \wh{\wt K}(t_2,x,v)=\wh{\wt K}(t_1+t_2, u,v) \ , \ee
as well as two other relations that can  be obtained from this one by differentiating with respect to $t_1$ or $t_2$.

\subsection{Small $t$  and small distance expansions}

The heat kernel $K(t,x,y)$ has a well-known small $t$-expansion from which follows the small $t$-expansion of $\wt K(t,x,y)$:
\be\label{Kasymp}
\wt K(t,x,y) \sim \frac{1}{4\pi t} e^{-\ell^2/4t} \Big[ a_0(x,y)+ t a_1(x,y) + t^2 a_2(x,y) +\ldots \Big] - \frac{e^{R_* t}}{A}
\ee
where $\ell^2\equiv\ell^2_A(x,y)$ is the geodesic distance squared between $x$ and $y$. For small $t$, the exponential forces $\ell^2$ to be small (of order $\sqrt{t}$) and we can use normal coordinates around $y$. Then 
\be\label{normalcoord}
\ell^2=(x-y)^2\equiv z^2  \quad \text{in normal coordinates}
\ee
and (see e.g.~the appendix of \cite{BF})
\ba
a_0(x,y)&=&1+\frac{R_*}{24} z^2 +\frac{R_*^2}{640} (z^2)^2 +\ldots \\
a_1(x,y) &=&\frac{7}{6} R_* +\frac{1}{20} R_*^2 z^2 +\ldots \\
a_2(x,y)&=& \frac{41}{60} R_*^2 +\ldots \ ,
\ea
as well as 
\be\label{grootexp}
\sqrt{g_*(x)}=\frac{1}{(a_0(x,y))^2}=1-\frac{R_*}{12} z^2+\frac{R_*^2}{480} (z^2)^2 +\ldots\ .
\ee
In particular, at coinciding points $y=x$,
\be\label{Ktildexx}
\wt K(t,x,x)\sim  \frac{1}{4\pi t} \Big[1 + \Big(\frac{7}{6}R_*-\frac{4\pi}{A}\Big) t + \Big( \frac{41}{60}R_*-\frac{4\pi}{A}\Big) R_* t^2 + \ldots\Big] 
\ee
is independent of the point $x$ since $R_*$ is constant. Then,  for any function $f$ that does not include the zero-mode, this implies that $\int \d^2 x \sqrt{g_*}\, \wt K(t,x,x) f(x)=0$. In particular,
\be\label{nolinearterm}
\int \d^2 x \sqrt{g_*}\, \wt K(t,x,x)\, \wt\f(x) = 0 \ ,
\ee
and the linear term in $S_{\rm measure}$ vanishes, as already anticipated.

To get the small $t$ expansion of $\wh{\wt K}(t,x,y)$ we use \eqref{KhatGint}. This yields
\be\label{Khattilsmallt}
\wh{\wt K}(t,x,y)=\wt G(x,y) -\frac{1}{4\pi}\sum_{k\ge 0} a_k(x,y) t^k E_{k+1}\Big(\frac{\ell^2}{4t}\Big) +  \int_0^t \d t' \, \frac{e^{R_* t'}}{A} \ ,
\ee
where the exponential integral functions $E_{n}$ are defined as
\be\label{Endef}
E_n(w)=\int_1^\infty \d u\, u^{-n} e^{- u w}\ ,
\ee
with asymptotic behaviours for small  $a$: $E_1(a)=-\g -\ln a + a+{\cal O}(a^2)$ and $E_2(a)=1+(\ln a+\g -1) a +{\cal O}(a^2)$.

We will be mostly interested in the case when $x$ is close to $y$ where one can use the small distance expansion of the Green's function.
One can show that the Green's function $\wt G(x,y)$ has the following expansion for $x$ close to $y$ (in normal coordinates around $y$):
\begin{multline}\label{Green1}
4\pi \wt G(x,y) \sim -\ln\frac{\m^2 z^2}{4} +4\pi \wt G_\zeta(y) -2\g + 2\pi z^i \del^i_y \wt G_\zeta(y) +z^2 \frac{\pi}{A}\\
-\frac{R_* z^2}{4} \left[ -\ln\frac{\m^2 z^2}{4} + 4\pi \wt G_\zeta(y) -2\g +\frac{7}{3} \right] \\
+\frac{1}{2} \left( z^i z^j-\frac{z^2}{2} \delta^{ij}\right) \del_x^i\del_x^j \wt C(x,y)\vert_{x=y} + \ldots \ .\hskip2.2cm
\end{multline}
This is obtained from first fixing the leading singularity so that for $x$ close to $y$ one has $\D_*^x \wt G(x,y) \sim \delta(x-y)$ (in normal coordinates around $y$, $\sqrt{g_*(y)}=1$), then adjusting the subleading terms so that $(\D_*^x-R_*) \wt G(x,y)=-\frac{1}{A}$ for $x\ne y$, and finally fixing the ``integration constants" in terms of $\wt G_\zeta$. In particular, $\wt C(x,y)$ is a symmetric function such that
\be
\wt C(y,y)= 4\pi \wt G_\zeta(y) +B \ ,
\ee
where $B$ is some constant that drops out. Upon inserting \eqref{Green1} into \eqref{Khattilsmallt} we get for small $t$ and small $z\equiv x-y$
\begin{multline}\label{Khattildexyexp}
\wh{\wt K}(t,x,y)=\frac{1}{4\pi} \Bigg\{
-\ln\frac{\m^2 z^2}{4} +4\pi \wt G_\zeta(y) -2\g + 2\pi z^i \del^i_y \wt G_\zeta(y)
-E_1\left( \frac{z^2}{4t}\right) \\
-\frac{z^2}{4} R_*\left[ -\ln\frac{\m^2 z^2}{4} + 4\pi \wt G_\zeta(y) -2\g +\frac{7}{3} 
+\frac{1}{6}E_1\left( \frac{z^2}{4t}\right)  \right]  +\frac{z^2}{4} \frac{4\pi}{A}\\
\hskip1.5cm + t\left[\frac{4\pi}{A}-\frac{7}{6} R_* E_2\left( \frac{z^2}{4t}\right)     \right]
+\frac{1}{2} \left( z^i z^j-\frac{z^2}{2} \delta^{ij}\right) \del_x^i\del_x^j \wt C(x,y)\vert_{x=y} \Bigg\}\\
+ {\cal O}(t^2, t z^2, z^3, z^3 \ln \m^2 z^2 ) \ . \hskip6.2cm
\end{multline}
As long as $t>0$, this has a smooth limit as $x\to y$ given by
\be\label{Khattildexxexp}
\wh{\wt K}(t,y,y)=\frac{1}{4\pi} \left[
-\ln\m^2 t +4\pi \wt G_\zeta(y) -\g 
+ t\left(\frac{4\pi}{A}-\frac{7}{6} R_*    \right)
\right]
+ {\cal O}(t^2) \ .
\ee
Note that the only term in $\wh{\wt K}(t,y,y)$ that depends on the position $y$ is $\wt G_\zeta(y)$.

For later use, let us also write the explicit form of $\wt K(t,x,y)$ using normal coordinates $z=x-y$ around $y\,$:
\be\label{Ktilexplicit}
\wt K(t,x,y)=\frac{e^{-z^2/4t}}{4\pi\, t}\left[ 1+\frac{1}{6}R_*\, t\left(\frac{z^2}{4t}  +7\right)+\ldots\right] -\frac{e^{R_* t}}{A} 
\ ,
\ee
as well as
\be\label{dsurdtK}
-\frac{\d}{\d t} \wt K(t,x,y)=\frac{e^{-z^2/4t}}{4\pi\, t^2}
\left[ \left( 1-\frac{z^2}{4t}\right) \left(1+\frac{R_*}{24}z^2\right) -\frac{7}{6} R_*  \frac{z^2}{4 } +\ldots
\right] +\frac{R_*}{A} e^{R_*t}\ .
\ee

\subsection{The regularized two-loop contributions}

We now give the regularized Feynman integrals corresponding to the two-loop diagrams \eqref{settingsun1}, \eqref{glasses1} and \eqref{figeight1}, as well as the measure diagram \eqref{measurediag}. We will write down the corresponding $I(t_i)$ as defined in \eqref{Feynmanreg}, subsequent multiplication with the $\vf(\a_i)$ and integration $\d\a_i$ being implicitly understood. One immediately gets
\begin{multline}\label{settingsun2}
I_{\diagramb}(t_1, t_2, t_3)
=\frac{4\pi}{\k^2} \int\d^2 x \sqrt{g_*(x)}\d^2 y \sqrt{g_*(y)}\,
\wh{\wt K}(t_1,x,y) \\
\times\Bigg[ \wh{\wt K}(t_2,x,y) (\D_*^x-\frac{2}{3}R_*)(\D_*^y-\frac{2}{3}R_*)\wh{\wt K}(t_3,x,y) \\
+ 2 \big[(\D_*^x-\frac{2}{3}R_*)\wh{\wt K}(t_2,x,y) \big](\D_*^y-\frac{2}{3}R_*)\wh{\wt K}(t_3,x,y) \Bigg]\ ,
\end{multline}
\begin{multline}\label{glasses2}
I_{\diagramd}(t_1, t_2, t_3)=\frac{2\pi}{\k^2} \int\d^2 x \sqrt{g_*(x)}\d^2 y \sqrt{g_*(y)}\\
\hskip3.cm\times\Bigg[  (\D_*^x-\frac{2}{3}R_*)\wh{\wt K}(t_1,x,x) + 2  (\D_*^x-\frac{2}{3}R_*)\wh{\wt K}(t_1,x,z)\big\vert_{z=x}\Bigg]\,  \wh{\wt K}(t_2,x,y) \\
\times
\Bigg[  (\D_*^y-\frac{2}{3}R_*)\wh{\wt K}(t_3,y,y) + 2  (\D_*^y-\frac{2}{3}R_*)\wh{\wt K}(t_3,y,z)\big\vert_{z=y}\Bigg] \ ,
\end{multline}
\begin{multline}\label{figeight2}
I_{\diagramc}(t_1,t_2)=-\frac{8\pi}{\k^2} \int\d^2 x \sqrt{g_*(x)}\,
\Bigg[ \frac{1}{4} \wh{\wt K}(t_1,x,x) (\D_*^x-2R_*)\wh{\wt K}(t_2,x,x) \\
\hskip3.cm+ \frac{1}{2} \Big[  (\D_*^x-2R_*) \big(\wh{\wt K}(t_1,x,z)\wh{\wt K}(t_2,x,z)\big)\Big]\Big\vert_{z=x}\\
+ 2 \wh{\wt K}(t_1,x,x) \big[ \D_*^x \wh{\wt K}(t_2,x,z)\big]\Big\vert_{z=x} \Bigg] \ ,
\end{multline}
\be\label{measurediag2}
\hskip-5.cm I_{\small\diagrammeasureloopsmall}(t_1,t_2)=\frac{4\pi}{\k^2} \int\d^2 x \sqrt{g_*(x)}\, \wt K(t_1,x,x) \wh{\wt K}(t_2,x,x) \ .
\ee
One should keep in mind that one can always symmetrize in the $t_i$ since only the symmetric part contributes.

\subsection{Scalings}

Before we further evaluate these integrals, it is useful to discuss some general features about their dependence on the area $A$.

The  eigenfunctions $\p_r$ of $D=\D_*-R_*$ are normalized as
$\int\d^2 x \sqrt{g_*}\, \p_r(x)\p_s(x)=\delta_{rs}$. It follows from \eqref{gstar} that $\p_r$ and $\l_r$ scale as
\be\label{lampsiAdep}
\l_r\equiv\l_r^A=\frac{A_0}{A} \,\l_r^{A_0} \quad , \quad \p_r\equiv\p_r^A=\sqrt{\frac{A_0}{A}} \,\p_r^{A_0} \ ,
\ee
where the quantities with a label $A_0$ are defined as the eigenfunctions and eigenvalues of $\D_0-R_0$ with respect to the metric $g_0$ of area $A_0$.
This immediately implies the scaling relations
\be\label{KKhatAdep}
{\wt K}_A(t,x,y)=\frac{A_0}{A}\ {\wt K}_{A_0}(\frac{A_0}{A}\, t,x,y)
 \quad , \quad \
\wh{\wt K}_A(t,x,y)=\wh{\wt K}_{A_0}(\frac{A_0}{A}\, t,x,y) \ .
\ee
The last relation implies in particular that $\wt G(x,y)=\wh{\wt K}(0,x,y)$ does not depend on $A$:
\be\label{GAindep}
\tilde G_A(x,y)=\tilde G_{A_0}(x,y) \ .
\ee
It then straightforwardly follows from \eqref{KKhatAdep} together with  \eqref{gstar}  that  all the integrals $I_n(t_i)$ as given in \eqref{settingsun2}-\eqref{measurediag2} satisfy
\be\label{IrAdep}
I_r[t_i, A]= I_r[\frac{A_0}{A}\, t_i, A_0] \ .
\ee
Of course, this is true only because these integrals are finite convergent integrals (for $t_i >0$). In particular, since $t_i=\a_i/\L^2$, we see that the $I_r$ cannot depend on $A$ and $\L^2$ separately but only on the combination $A\L^2$. On the other hand, the small $t$ short-distance expansion given above for  $\wh{\wt K}$ also depends on an arbitrary scale $\m$ introduced when defining $\wt G_\zeta$ in \eqref{Gzetadef}, so one might wonder whether an additional area-dependence of the form $\m^2 A$ could occur. However, this is not the case. Indeed, in \eqref{Gzetadef} the scale $\mu$ appears in the combination $\ln [\ell^2_A(x,y) \m^2]$ where $\ell_A(x,y)$ is the geodesic distance between $x$ and $y$ computed with the metric of area $A$. Thus  $\ln[ \ell^2_A(x,y) \m^2] = \ln A \m^2 + \ldots$ where $+\ldots$ refers to terms that do not depend on $A$ or $\m$. It follows that $4\pi \wt G_\zeta - \ln A \m^2$ does not depend on $\m$. Since also, as we will explain shortly,
\be\label{GzetaAscale}
\wt G^A_\zeta = \wt G^{A_0}_\zeta + \frac{1}{4\pi} \ln\frac{A}{A_0} \ , 
\ee
it follows that
\be
4\pi \wt G_\zeta^A-\ln \frac{\ell^2_A(x,y) \m^2}{4}=4\pi \wt G^{A_0}_\zeta -\ln A_0\m^2 +\ldots \ ,
\ee
depends neither on $A$ nor on $\m$. Since $\wt G_\zeta$ and $\m$ only ever appear in this combination in $\wh{\wt K}$ it is clear that there is no real  $\m$ dependence in the end. Thus
\be\label{Irdeps}
I_r[\frac{\a_i}{\L^2}, A]=f_r[\L^2 A, \frac{\a_i}{\a_j}] \ . 
\ee
For example, a term $\big( \ln \L^2 A\big)^2$ will appear as
\be
\big( \ln \L^2 A\big)^2=\left( \ln \frac{\L^2}{\m^2} + \ln\frac{A}{A_0} + \ln \m^2 A_0 \right)^2 =\left( \ln\frac{A}{A_0} \right)^2 + 2\ln \frac{\L^2}{\m^2} \  \ln\frac{A}{A_0}  + \ldots
\ee
where $+\ldots$ refers to terms independent of the area $A$. This structure will be indeed explicitly observed below. Similarly, a term $A\L^2 \ln \m^2A$ must be accompanied by a term $A \L^2 \ln \frac{\L^2}{\m^2}$. Again, this will be explicitly the case.

The relation \eqref{GzetaAscale} that gives the area dependence of
$\wt G^A_\zeta$ will play a most important role below, since it is through this relation that the $\ln\frac{A}{A_0}$ terms appear in the logarithm of the partition function. Let us prove it. To begin with, $\wt G_\zeta(y)$ is defined in terms of the spectral $\zeta$-function (without zero-mode)
\be\label{zetatilde}
\wt\zeta(s,x,y)=\sum_{r>0}\frac{\p_r(x)\p_r(y)}{\l_r^s} \ .
\ee
Clearly, $\wt\zeta(1,x,y)=\wt G(x,y)$ is singular as $x\to y$. For $s\ne 1$,  $\wt\zeta(s,x,y)$  also provides a regularization of the propagator. More precisely,  $\wt \zeta(s,x,x)$ is a meromorphic function with a pole at $s=1$.  The spectral $\zeta$-function is related  to our heat kernel ${\wt K}(t,x,y)$ by the Mellin transformation $\wt\zeta(s,x,y)=\frac{1}{\G(s)}\int_0^\infty \d t\, t^{s-1} \wt K(t,x,y)$. From this relation it is not difficult to show that the residue of the pole of $\wt\zeta(s,x,x)$ at $s=1$ is $\frac{a_0(x,x)}{4\pi}=\frac{1}{4\pi}$. Then one can define, cf \cite{BF}
\be\label{Gzetazetadef}
\wt G_\zeta(x)=\lim_{s\to 1}\left[ \m^{2(s-1)} \wt\zeta(s,x,x)-\frac{1}{4\pi(s-1)}\right] \ ,
\ee
which is equivalent to saying
\be\label{Gzeta2}
\m^{2(s-1)}\wt\zeta(s,x,x)=\frac{1}{4\pi(s-1)} + \wt G_\zeta(x) + {\cal O}(s-1) \ .
\ee
From \eqref{zetatilde} and \eqref{lampsiAdep} it follows that
\be\label{zetascale}
\wt \zeta_A(s,x,x)=\left(\frac{A}{A_0}\right)^{s-1} \wt\zeta_{A_0}(s,x,x) \ .
\ee
Inserting this into \eqref{Gzetazetadef} or \eqref{Gzeta2} for $\wt G_\zeta^A$ and rewriting the r.h.s. in terms of $\wt G_\zeta^{A_0}$ yields the desired relation \eqref{GzetaAscale}.
It remains to show that $\wt G_\zeta$ defined by  \eqref{Gzetazetadef} is exactly the same quantity as the one defined by \eqref{Gzetadef} and that appears in the expansion \eqref{Khattildexyexp} of $\wh{\wt K}$. This is done as follows. By the Mellin transformation between $\wt\zeta(s,x,y)$ and the heat kernel, the singularity of the former is related to the small $t$ asymptotics of the latter and one sees that
\be\label{zetaR}
\wt \zeta_R(s,x,y)=\wt\zeta(s,x,y)-\frac{1}{\G(s)}\int_0^{1/\m^2} \frac{d t}{4\pi t^2} t^s a_0(x,y) e^{-\ell^2(x,y)/4t}
\ee
is smooth for $s\to 1$ and for $y\to x$. Taking first the limit $y\to x$ (for $s>1$) and then $s\to 1$ yields $\wt G_\zeta(x)$. On the other hand, taking first $s\to 1$ yields 
$\wt G(x,y)-\frac{1}{4\pi}a_0(x,y) E_1(\frac{\m^2\ell^2(x,y)}{4})$ and then letting $y\to x$ yields the relation \eqref{Gzetadef}.

\section{Evaluating the regularized two-loop integrals}

We first rewrite the integrals  $I_n(t_i)$ as given in \eqref{settingsun2}-\eqref{measurediag2} by using the relations \eqref{KKhatrel} to convert as many $\wh{\wt K}$ into $\wt K$ as possible. We also use the fact that the only $x$-dependent piece in $\wh{\wt K}(t,x,x)$ is $\wt G_\zeta(x)$ (at least for small $t$) and that ${\wt K}(t,x,x) $ does not depend on $x$. Hence, due to the absence of a zero-mode,
\ba\label{simplfyrel1}
\int\d^2 x \sqrt{g_*(x)}\, \wh{\wt K}(t_1,x,x) \wh{\wt K}(t_2,x,y)&=&\int\d^2 x \sqrt{g_*(x)}\, \wt G_\zeta(x) \wh{\wt K}(t_2,x,y) \ ,\\
\int\d^2 x \sqrt{g_*(x)}\, {\wt K}(t_1,x,x) \wh{\wt K}(t_2,x,y)&=& 0 \ ,
\ea
and similarly with $\wh{\wt K}(t_2,x,y)$ replaced by ${\wt K}(t_2,x,y)$. As already emphasized, since the $I_n(t_i)$ will be multiplied by $\prod_i \int \d\a_i \vf(\a_i)$,  all $t_i$ are effectively symmetrized.  Thus, we consider two expressions as identical if they only differ by a permutation of the $t_i$. Finally, $\nabla_i^x \wh{\wt K}(t,x,x)= 2 \nabla^x_i \wh{\wt K}(t,x,z)\big\vert_{z=x}$ so that
\begin{multline}
\D^x \Big(\wh{\wt K}(t_1,x,z)\wh{\wt K}(t_2,x,z)\Big)\Big\vert_{z=x}
= 2 \big(\D^x \wh{\wt K}(t_1,x,z)\big)\big\vert_{z=x}\wh{\wt K}(t_2,x,x)\\
-\frac{1}{2} g^{ij} \nabla_i^x \wh{\wt K}(t,x,x) \nabla_j^x \wh{\wt K}(t,x,x) \ .
\end{multline}
It is then possible to express  \eqref{settingsun2}-\eqref{measurediag2} 
as
\ba\label{IdiagJi}
I_{\diagramb}&=& J_1+J_3+J_4 + J_6 \ , \nonumber \\
I_{\diagramd}&=&J_2+J_5+J_7 \ ,  \nonumber \\
I_{\diagramc}&=& - 3 J_8-3 J_9 \ ,\nonumber  \\
I_{\diagrammeasureloopsmall}&=& \frac{1}{2} J_8 
\ea
in terms of the following basic integrals (the notation $\int\d x$ is short-hand for $\int \d^2 x \sqrt{g_*(x)}$):
\ba\label{Jidefs}
J_1&=&\frac{4\pi}{\k^2}\int \d x \d y \ \wh{\wt K}(t_1,x,y) 
\left(-\frac{\d}{\d t_2} \wt K(t_2,x,y)\right) \wh{\wt K}(t_3,x,y)\ , 
\\
J_2&=&\frac{2\pi}{\k^2}\int \d x \d y \ \wt G_\zeta(x)
\left(-\frac{\d}{\d t_2} \wt K(t_2,x,y)\right) \wt G_\zeta(y)\ , 
\\
J_3&=&\frac{8\pi}{\k^2}\int \d x \d y \ \wh{\wt K}(t_1,x,y) 
\wt K(t_2,x,y) {\wt K}(t_3,x,y) \ , 
\\
J_4&=&\frac{8\pi}{\k^2}R_*\int \d x \d y \ \wh{\wt K}(t_1,x,y) 
\wt K(t_2,x,y)\wh{\wt K}(t_3,x,y)  \ , 
\\
J_5&=&\frac{4\pi}{\k^2}R_*\int \d x \d y\  \wt G_\zeta(x){\wt K}(t_2,x,y) \wt G_\zeta(y)\ , 
\\
%
J_6&=&\frac{4\pi}{3\k^2}R_*^2 \int\d x \d y\  \wh{\wt K}(t_1,x,y)
\wh{\wt K}(t_2,x,y)\wh{\wt K}(t_3,x,y)\ , 
\\
J_7&=&\frac{2\pi}{\k^2}R_*^2\int \d x \d y\  \wt G_\zeta(x)\wh{\wt K}(t_2,x,y) \wt G_\zeta(y)\ , \\
J_8&=&\frac{8\pi}{\k^2}\int \d y \ \wh{\wt K}(t_1,y,y) {\wt K}(t_2,y,y) \ ,
\\
J_9&=&\frac{4\pi }{\k^2}R_*\int \d y \ \wh{\wt K}(t_1,y,y) \wh{\wt K}(t_2,y,y)\ , 
\ea
%
We now evaluate these integrals $J_i$ one by one. Of course, we must keep in mind that we are free to drop all terms that vanish as $\L\to\infty$, i.e. terms that overall are ${\cal O}(t)$ (e.g. a term $\frac{t_1 t_2}{t_3}$). Also any area-independent finite terms are without interest since they drop out when computing $Z[A]/Z[A_0]$. We do, however, keep the area-independent diverging terms in order to check that, in the end, they only show up in the combinations allowed by the above scaling argument, see \eqref{Irdeps}.

\subsection{The integrals  $J_2,\ J_5,\ J_7$}

The integrals  $J_2,\ J_5,\ J_7$ are needed to compute $I_{\diagramd}$.
We begin with $J_7$. This integral has a finite limit as $\L\to\infty$, i.e. as $t_2\to 0$. Indeed, in this limit one simply replaces $\wh{\wt K}(t_2,x,y)$ by $\wt G(x,y)$ which has an integrable logarithmic short-distance singularity. Thus $J_7=\frac{2\pi}{\k^2}R_*^2\int \d x \d y\  \wt G_\zeta(x) \wt G(x,y) \wt G_\zeta(y)+{\cal O}(1/\L)$. Using \eqref{GzetaAscale}, together with $\int\d x \, \wt G(x,y)=\int\d y\,  \wt G(x,y)=0$,  one further finds that 
\be\label{J7}
J_7=\frac{2\pi}{\k^2}R_*^2\int \d x \d y\  \wt G^{A_0}_\zeta(x)\wt G(x,y) \wt G^{A_0}_\zeta(y) +{\cal O}(1/\L)=c_7 +{\cal O}(1/\L) \ ,
\ee
where we denote by $c_i$ various $A$-independent finite constants. Indeed, $\wt G_\zeta^{A_0}$ and $\wt G$ do not depend on $A$, $\int\d x\d y\equiv \int\d^2 x \sqrt{g_*(x)}\,\d^2 y \sqrt{g_*(y)}$ scales as $A^2$ and $R_*^2\sim \frac{1}{A^2}$ so that the first term obviously is an $A$-independent constant.

Next, we show in some detail how to evaluate $J_5$. We write $\wt K(t_2,x,y)$\break$=K(t_2,x,y)-\frac{e^{R_* t_2}}{A}$ and use the fact that for small $t$ the heat kernel $K(t_2,x,y)$ vanishes exponentially unless $\ell(x,y)^2$ is of order $t_2$ or less. Thus we can use normal coordinates $z$ around $y$ and use the expressions \eqref{Ktilexplicit} and \eqref{grootexp}
\begin{multline}\label{J5}
\hskip-2.mmJ_5=\frac{4\pi}{\k^2}R_*\hskip-1.5mm\int\hskip-1.5mm \d y \,\d^2 z \Big( 1-\frac{R_*}{12}z^2 +\ldots\Big)  \wt G_\zeta(y+z) \frac{e^{-z^2/4t_2}}{4\pi\, t_2}\left[ 1+\frac{R_*}{6}\, t_2\left(\frac{z^2}{4t_2}  +7\right)\hskip-1.mm+\ldots\right] \hskip-1.mm  \wt G_\zeta(y)\\
-\frac{4\pi}{\k^2}\frac{R_*e^{R_* t_2}}{A}\left( \int\d y\, \wt G_\zeta(y)\right)^2 \ .
\end{multline}
Note that we also keep the subleading terms $\sim t_2$ since below we will obtain $J_2$ simply by taking $-\frac{\d}{\d t_2}$ of  $J_5(t_2)$ and dropping a factor of $R_*$. These subleading terms in $J_5$ will yield the finite terms of $J_2$. Now $\wt G_\zeta(y+z)$ is a smooth function and can be expanded in a series around $z=0$. Performing the Gaussian integrals then yields
\begin{multline}\label{J5-2}
J_5=\frac{4\pi}{\k^2}R_*\hskip-1.mm\int\hskip-1.mm \d y  \left[ \wt G_\zeta^2(y) (1+R_* t_2) 
-t_2 \wt G_\zeta(y)\D_* \wt G_\zeta(y)   \right]
-\frac{4\pi}{\k^2}\frac{R_*e^{R_* t_2}}{A}\left( \int\d y\, \wt G_\zeta(y)\right)^2 \\
+{\cal O}(t_2^2) \ .
\end{multline}
Thus
\be\label{J5-3}
J_5=\frac{4\pi}{\k^2}R_*\int \d y\, \wt G_\zeta^2(y) 
-\frac{4\pi}{\k^2}\frac{R_*}{A}\left( \int\d y\, \wt G_\zeta(y)\right)^2
+{\cal O}(1/\L^2) =c_5 +{\cal O}(1/\L^2)\ ,
\ee
where we used again \eqref{GzetaAscale} to show that the explicitly written finite terms do not depend on $A$.  Taking $-\frac{\d}{\d t_2}$ of \eqref{J5-2} and dropping a factor of $R_*$ we also get
\ba\label{J2}
J_2&=&\frac{4\pi}{\k^2} \int \d y\, \left[ R_*\wt G_\zeta^2(y) -\wt G_\zeta(y)\D_* \wt G_\zeta(y)  \right]-\frac{4\pi}{\k^2}\frac{R_*}{A}\left( \int\d y\, \wt G_\zeta(y)\right)^2
+{\cal O}(1/\L^2)\nonumber \\
&=& c_2 + {\cal O}(1/\L^2)  \ ,
\ea
since the first plus the third term are the same as in \eqref{J5-3}, and for the second term \eqref{GzetaAscale} implies that $\int \wt G_\zeta\D_* \wt G_\zeta= \int \wt G_\zeta^{A_0}\D_* \wt G_\zeta^{A_0}$ which is also independent of $A$.
Thus we conclude that
\be\label{glassesresult}
I_{\diagramd}=J_2+J_5+J_7=c_2+c_5+c_7+{\cal O}(1/\L^2)\ .
\ee
It is satisfying that this diagram does not give any non-trivial contribution. Indeed, we expect that the propagator connecting the two loops should carry zero ``momentum" and, since  no zero-mode is present, we expect to obtain zero. The absence of a zero-mode, of course, was the reason that the $\wh{\wt K}(t_i,x,x)$ could be replaced by $\wt G_\zeta(x)$, and similarly for $y$. However, on a non-trivial manifold $\wt G_\zeta(x)$ is not constant, and the overall contribution does not need to vanish. We also note that the integrals $J_2$, $J_5$ and $J_7$ are all finite and area-independent, in agreement with our general argument that $A$ can only appear in the combination $A\L^2$: only divergent integrals can give an area dependence.

\subsection{The integrals  $J_8$ and $J_9$}

The integrals  $J_8$ and $J_9$ are needed to compute $I_{\diagramc}$ and $I_{\diagrammeasureloopsmall}$.
They are single integrals over the manifold. It is then straightforward to multiply the asymptotic expansions \eqref{Khattildexxexp} and \eqref{Ktilexplicit} for $x=y$ to get the relevant expansion of the integrands. We find
\begin{multline}\label{J8-1}
J_8= \frac{1}{2\pi\k^2}
\left( \frac{\L^2}{\a_2}+\frac{7}{6}R_* -\frac{4\pi}{A}\right)
\int\d y\, \left[ 4\pi\wt G_\zeta(y) + \ln\frac{\L^2}{\m^2 \a_1}-\g\right] +c_8+{\cal O}(1/\L^2).
\end{multline}
Using once more \eqref{GzetaAscale}, it is not difficult to see that the terms combine so that $A$ and $\L^2$ always appear in the combination $A\L^2$ and $\m^2$ in the combination $A_0\m^2$. Next, 
\begin{multline}\label{J9-1}
J_9=\frac{R_*}{4\pi \k^2} \int\d y\, 
\left[4\pi\wt G_\zeta(y) +\ln\frac{\L^2}{\m^2\a_1}-\g\right]
\left[4\pi\wt G_\zeta(y) +\ln\frac{\L^2}{\m^2\a_2}-\g\right] +{\cal O}(1/\L^2)\ .
\end{multline}

\subsection{The integrals  $J_1$, $J_3$, $J_4$ and $J_6$}

These are the  integrals  needed to compute $I_{\diagramb}$\ .
We begin with $J_6$. If we replace in this integral each of the $\wh{\wt K}(t_i,x,y)$ by the unregulated Green's function $\wt G(x,y)$ we get a converging finite integral, since the short-distance singularity $\sim (\ln \ell(x,y))^3$ is integrable. Now, $\wt G(x,y)$ does not depend on the area and, thus 
\be\label{J6}
J_6=c_6 +{\cal O}(1/\L^2) \ .
\ee
It remains to compute the integrals $J_1$, $J_3$, $J_4$. They all involve at least one $\wt K(t_i,x,y)$ or $\frac{d}{\d t_2}\wt K(t_2,x,y)$. Except for the zero-mode contribution, this factor is exponentially small unless $\ell^2(x,y)$ is of order $t_i$ or less, allowing us to use normal coordinates $z=x-y$ and do the various integrations over $z$ explicitly. For convenience, we have listed the relevant integrals in the apendix \ref{integrals}.
We always let
\be\label{Jisep}
J_i=J_i^{(1)}-J_i^{(2)} \ , \quad i=1,3,4 \ ,
\ee
where $J_i^{(1)}$ refers to the parts where all the $\wt K$ are replaced by $K$ and 
$J_i^{(2)}$ refers to the remaining parts where at least one $\wt K$ is replaced by the subtracted zero-mode. The subtracted zero-mode part of $\frac{d}{\d t_2}\wt K(t_2,x,y)$ is $-\frac{R_*}{A} +{\cal O}(t_2)$, so that $J_1^{(2)}=-\frac{4\pi R_*}{\k^2 A} \int\d x\, \d y\, \wh{\wt K}(t_1,x,y)\wh{\wt K}(t_3,x,y)$ $+{\cal O}(t_2)$.
Using the by now familiar argument of replacing the $\wh{\wt K}(t_i,x,y)$ by $\wt G(x,y)$ we see that $J_1^{(2)}$ is a finite area-independent constant, up to terms  ${\cal O}(1/\L^2)$. For $J_4^{(2)}$ the argument works exactly the same way. Thus
\be\label{J12}
J_1^{(2)}=c_1^{(2)}+{\cal O}(1/\L^2) 
\quad , \quad
J_4^{(2)}=c_4^{(2)}+{\cal O}(1/\L^2) \ .
\ee
For $J_3^{(2)}$ one gets two contributions that contribute equally (upon symmetrizing in $t_2$ and $t_3$):
\ba\label{J32}
J_3^{(2)}&=&\frac{16\pi}{\k^2}\frac{e^{R_*t_3}}{A}\int\d x \d y \, \wh{\wt K}(t_1,x,y) K(t_2,x,y)  \nonumber \\
&=&\frac{4}{\k^2}\, \frac{1}{A} \int\d y\, 
\left[4\pi\wt G_\zeta(y) +\ln\frac{\L^2}{\m^2(\a_1+\a_2)}-\g\right]+{\cal O}(1/\L^2) \ .
\ea

It remains to compute the $J_i^{(1)}$. The simplest is $J_4^{(1)}$ where it is easy to see that we only need to keep the leading term in the small $t_i$ expansion of $K(t_2,x,y)$, cf \eqref{Ktilexplicit}, and that we can drop all terms ${\cal O}(t_i)$ or ${\cal O}(z^j)$ in the expansion of $\wh{\wt K}$, cf \eqref{Khattildexyexp}, or of $\sqrt{g_*}$. The relevant integrals that multiply terms involving either $\wt G_\zeta$ and/or $\ln\m^2 t_2$ are all listed in the appendix \ref{integrals}. Other integrals like $\int\d^2 \tilde z e^{-\tilde z^2} E_1(\tilde z^2 \frac{t_2}{t_1})E_1(\tilde z^2 \frac{t_2}{t_3})$ only contribute finite constants that only depend on ratios of the $\a_j$. We get
\be\label{J41}
J_4^{(1)}=\frac{R_*}{2\pi\k^2} \int \d y \left[4\pi \wt G_\zeta(y) +\ln\frac{\L^2}{\m^2(\a_1+\a_2)}-\g\right]^2 + c_4^{(1)}(\a_i)+{\cal O}(1/\L^2) \ .
\ee
Next, we compute $J_1^{(1)}$. With respect to the previous computation, here $-\frac{\d}{\d t_2} K(t_2,x,y)$ involves one more factor of $\frac{1}{t_2}$, so that in the expansions of $\wh{\wt K}(t_i,x,y)$ one has also to keep the terms ${\cal O}(t_i)$ or ${\cal O}(z^iz^j)$. The computation then is quite lengthy but straightforward. Note that a term $\int\d y \del_i\wt G_\zeta \del^i \wt G_\zeta$ appears. Integrating by parts, this equals $\int \wt G_\zeta \D_* \wt G_\zeta$ which was shown above to be an $A$-independent constant. Upon using the symmetry under exchange of the $\a_i$, the result then is
\begin{multline}\label{J11}
J_1^{(1)}=\frac{R_*}{4\pi\k^2} \int\d y\, 
\left[ 4\pi \wt G_\zeta(y)+\ln\frac{\L^2}{\m^2(\a_1+\a_2)}-\g+\frac{1}{6} -\frac{\a_1\a_2}{(\a_1+\a_2)^2}\right]^2\\
+\frac{1}{2\pi\k^2} \Big(\frac{\L^2}{\a_1+\a_2}-\frac{4\pi}{A}\Big)\hskip-1.mm \int\hskip-1.mm\d y
\left[ 4\pi \wt G_\zeta(y)+\ln\frac{\L^2}{\m^2\a_2}-2\g\right]\\
+\frac{1}{2\pi\k^2}\left(\L^2 A \,C_1^{(1)}(\a_i)  +c_1^{(1)}(\a_i)\right)+{\cal O}(1/\L^2)\ .
\hskip3.cm
\end{multline}
Note that one might have expected a leading singularity $\sim A\L^2 \left(\ln \frac{\L^2}{\m^2} \right)^2$, due to the leading $\frac{1}{t_2^2}$ singularity of $\frac{\d}{\d t_2}\wt K(t_2,x,y)$, see \eqref{dsurdtK} multiplying the $\ln \frac{\L^2}{\m^2}$ singularities from each of the two $\wh{\wt K}$. However, this leading term is multiplied by $\int \d^2\wt z e^{-\wt z^2}(1-\wt z^2)$ which vanishes. Thus the leading singularity is $\sim A\L^2 \ln  \frac{\L^2}{\m^2}$ as expected from naive power counting for the present two-loop diagrams.

Last, to compute $J_3^{(1)}$, we observe that from \eqref{grootexp} and \eqref{Ktilexplicit} we get
\be\label{grootKK}
\sqrt{g_*(x )}\, K(t_2,x,y)K(t_3,x,y)=\frac{e^{-z^2/4T}}{(4\pi)^2 (t_2+t_3)T}
\left[1+\frac{7R_*}{6}(t_2+t_3) +\ldots\right] \ ,
\ee
where $T=\frac{t_2 t_3}{t_2+t_3}$. Thus in $\wh{\wt K}(t_1,x,y)$ one has again to keep the terms ${\cal O}(t_1)$ or ${\cal O}(z^iz^j)$. Doing the integrals over $z$ results in
\begin{multline}
J_3^{(1)}=\frac{1}{2\pi \k^2} \frac{\L^2}{\a_2+\a_3} \int \d y\
\left[ 4\pi \wt G_\zeta(y)+\ln\frac{\L^2}{\m^2}-\g+\ln\frac{\a_2+\a_3}{\a_1\a_2+\a_1\a_3+\a_2\a_3}\right] \\
+\frac{R_*}{2\pi \k^2} \left( \frac{7}{6}-\frac{\a_2\a_3}{(\a_2+\a_3)^2}\right) 
\hskip-1.mm\int\hskip-1.mm\d y
\left[ 4\pi \wt G_\zeta(y)+\ln\frac{\L^2}{\m^2}\right] 
+c_3^{(1)}(\a_i)+{\cal O}(1/\L^2)\ .
\end{multline}

\subsection{Summing the two-loop and measure diagrams}

We have already seen, cf \eqref{glassesresult} that $I_{\diagramd}=c_2+c_5+c_7+{\cal O}(1/\L^2)$. For the remaining three diagrams we have from \eqref{IdiagJi}
\ba\label{Isum}
I_{\diagramb}+I_{\diagramc}+I_{\diagrammeasureloopsmall}
&=& J_1+J_3+J_4 + J_6-\frac{5}{2} J_8 -3 J_9 \nonumber\\
&=&J_1^{(1)} +J_3^{(1)}-J_3^{(2)}+J_4^{(1)}  -\frac{5}{2} J_8 -3 J_9 + c + {\cal O}(1/\L^2)\ .
\ea
Inserting our above results for the $J_i$, we find for the logarithm of the partition function
\begin{multline}\label{Isum2}
W^{(2)}[A]\big\vert_{\rm loops}=I_{\diagramb}+I_{\diagramc}+I_{\diagrammeasureloopsmall}+I_{\diagramd}
\\
=\frac{1}{2\pi\k^2} 
\left\{ R_* \left( \frac{3}{2}\ln\frac{\a_1\a_2}{(\a_1+\a_2)^2 } -2\frac{\a_1\a_2}{(\a_1+\a_2)^2}-\frac{19}{12}\right)
+\frac{2\L^2}{\a_1+\a_2}-\frac{5\L^2}{2\a_1} -\frac{2\pi}{A}\right\}\\
 \times\int\d y \left[4\pi\wt G_\zeta(y) +\ln\frac{\L^2}{\m^2}-\g\right]\ \ 
 +\ \frac{\L^2 A}{2\pi\k^2} \, C(\a_i) + c(\a_i) +{\cal O}(1/\L^2)  \ ,
\end{multline}
where $C(\a_i)$ and $c(\a_i)$ are finite area-independent but regulator-dependent ``constants". $C(\a_i)$ is given by (up to permutations and/or symmetrizations of the $\a_i$) 
\begin{multline}\label{Calpha}
C(\a_i)=\frac{1}{\a_1+\a_2} \left[ \frac{1}{2}\ln\frac{(\a_1+\a_2)^2}{\a_1\a_2} -\ln(\a_1\a_2+\a_1\a_3+\a_2\a_3)-\g\right] \\
+\frac{5}{4}\left[ \frac{\ln\a_1}{\a_2}+\frac{\ln\a_2}{\a_1}\right] + C_1^{(1)}(\a_i)\ .
\hskip5.cm 
\end{multline}
To explicitly display the area dependence, we use \eqref{GzetaAscale} to rewrite
\be\label{Adep}
\int\d y \left[4\pi\wt G_\zeta(y) +\ln\frac{\L^2}{\m^2}-\g\right]
=A \left[ \ln\frac{A\L^2}{A_0\m^2} +\frac{4\pi}{A_0}\int \d^2 y\,\sqrt{g_0}\, \wt G_\zeta^{A_0}(y)-\g\right] \ .
\ee
Quite remarkably, the individual integrals, $J_1$, $J_4$ and $J_9$ all involve the square of the expression in square brackets and, hence, contain $(\ln A \L^2)^2$ divergences. However, these integrals appear in exactly the right combination such that these $(\ln A \L^2)^2$-terms cancel in $W^{(2)}$.

We now define
\ba\label{alphafcts}
F(\a_i)&=&8\pi(1-h) \left( \frac{3}{2}\ln\frac{\a_1\a_2}{(\a_1+\a_2)^2 } -2\frac{\a_1\a_2}{(\a_1+\a_2)^2}-\frac{19}{12}\right) -2\pi \ ,\nonumber\\
H(\a_i)&=&\frac{2}{\a_1+\a_2}-\frac{5}{2\a_1} \ ,\nonumber\\
G_0&=&\frac{4\pi}{A_0}\int \d^2 y\,\sqrt{g_0}\, \wt G_\zeta^{A_0}(y)-\g \ ,
\ea
so that
\begin{multline}\label{Isum3}
W^{(2)}[A]\big\vert_{\rm loops}
=\frac{1}{2\pi\k^2} \left[\ln\frac{A\L^2}{A_0\m^2} +G_0\right] \Big[ F(\a_i)+A\L^2 H(\a_i)\Big]
 +\ \frac{\L^2 A}{2\pi\k^2} \, C(\a_i)\\
  + c(\a_i) +{\cal O}(1/\L^2) \ .\hskip7.5cm
 \end{multline}
This is indeed of the form \eqref{lnZAform} without the $(\ln A \L^2)^2$-term.
If we normalize with respect to $A_0$, i.e we compute $\ln\frac{Z[A]}{Z[A_0]}$, we get
\begin{multline}\label{Isum4}
W^{(2)}[A]\big\vert_{\rm loops}-W^{(2)}[A_0]\big\vert_{\rm loops} 
=\frac{1}{2\pi\k^2} \Big(F(\a_i)+A\L^2 H(\a_i)\Big)\,  \ln\frac{A}{A_0} \\
 +\ \frac{ (A-A_0)\L^2}{2\pi\k^2} \, \left[C(\a_i) + G_0 H(\a_i)+ H(\a_i) \ln\frac{\L^2}{\m^2}\right]
\ +{\cal O}(1/\L^2)\ .
\end{multline}
The last line corresponds to a cosmological constant term, and we may always add  a corresponding counterterm to the quantum gravity action to cancel this term.
The first line displays an $\ln\frac{A}{A_0}$-term, as expected, although with a rather complicated, regulator dependent coefficient, as well as a term $\sim A\L^2 \ln\frac{A}{A_0}$. The latter term is certainly not expected to occur. It is non-local, and it can  not be cancelled by any local {\it two-loop} counterterm.
At this point it is useful to mention that the KPZ result is, see \eqref{gammastrexp},
\be\label{KPZres}
\left[W^{(2)}[A]-W^{(2)}[A_0] \right]\Big\vert_{{\rm KPZ}}=\frac{2}{\kappa^2}(1-h) \ln\frac{A}{A_0} - \m_c^2 (A-A_0)\ .
\ee

Clearly, as in any quantum field theory, the contributions at order $\k^{-2}$ do not only come from two-loop diagrams, but also from  tree and one-loop diagrams involving counterterms. Of course, there is no ``vacuum tree-diagram" and the only ``tree" contribution is the cosmological constant counterterm we already mentioned. However, the various $n$-point functions at one loop (that are $\sim \k^{-n}$) are also divergent quantities and, to make them finite, one has to introduce various ``one-loop" counterterm $n$-point vertices that will be of order $\k^{-n}$. In particular, there will be a quadratic counterterm vertex needed to make the renormalized propagator finite, and the one-loop diagram made with this counterterm vertex will contribute at order $\k^{-2}$ to the partition function.
Thus, our next task is to determine this counterterm. This will involve the one-loop computation of the 2-point function.


\section{Counterterms and one-loop computation of the two-point function}

\subsection{Generalities and order $1/\k^2$ contributions from the counterterms}

Any divergence in $W[A]\big\vert_{\rm loops}$ that is simply proportional to the area $A$ (and hence also to $\L^2$) can be cancelled by a  counterterm of the form  $\int\d^2 x \sqrt{g_*} \L^2 \left( c_i \ln\frac{\L^2}{\m^2} +c_i\right)$, referred to as cosmological constant. Also, any $A$-independent constants drop out when computing $\ln\frac{Z[A]}{Z[A_0]}=W[A]-W[A_0]$ and are thus irrelevant.
However, the divergences $\sim \L^2 \int\d y\, 4\pi \wt G_\zeta= A\L^2 (\ln\frac{A}{A_0}+c)$ are {\it non-local} divergences. Of course, they cannot be cancelled by any local two-loop counterterm. However, they can be cancelled by one-loop diagrams involving local (one-loop) counterterm vertices. The same thing happens on a four-dimensional curved manifold  with non-derivative $\f^3$ and $\f^4$ couplings \cite{BF}.
At present, the only such one-loop contribution to the partition function comes from a diagram having a single propagator connecting the two legs of the conterterm vertex that itself is $\sim \frac{1}{\k^2}$. 
This is very similar to the diagram coming from the measure and can thus be seen as renormalizing the measure. In particular, the measure resulted in a vertex $\sim\int \d x \, \wt K(t,x,x) \wt\f^2(x)$ where $\wt K(t,x,x)=\frac{\L^2}{4\pi\a}+\frac{7}{24\pi}R_* -\frac{1}{A} +\ldots$ is a constant. Thus any counterterm that looks like an (unwanted) mass renormalization can actually be interpreted as a renormalization of the measure. We thus allow a counterterm action of the form
\be\label{Sct}
S_{\rm ct}= \frac{8\pi}{\k^2}\int\d^2 x \sqrt{g_*} \left[ \frac{c_\f}{2} \wt \f (\D_*-R_*)\wt \f +\frac{c_R}{2} R_* \wt\f^2 + \frac{c_m}{2} \wt\f^2 \right] \ .
\ee
Note that a linear counterterm automatically vanishes since $\wt\f$ has no zero-mode, and higher counterterm vertices only contribute within two-loop vacuum diagrams that are $\sim \frac{1}{\k^4}$.
The counterterm coefficients obviously can depend on the cutoff $\L$ and on the regularization functions $\vf(\a)$. They also have an expansion in powers of $\frac{1}{\k}$. Due to the explicit factor of $\frac{1}{\k^2}$ in front of the counterterm action, we will only be interested here in the lowest order, $\k$-independent pieces. Note that in terms of the original $\wh\f$ they are of order $\k^0\wh\f^2$, consistent with the fact that these terms originate at one loop.
The dependence on $\L$ is again dictated by dimensional considerations: $c_\f$ and $c_R$ must be dimensionless and will correspond to at most logarithmic divergences, while $c_m$ has dimension of $\L^2$ and corresponds to an at most quadratic (times a log) divergence. With $\int \d \a_i \vf(\a_i)$ implicitly understood\footnote{
This means that if any given $c(\a_i)$ depends on any given number of $\a_i, \ i=1,\ldots r$, one should really think of it as $c[\vf]=\int_0^\infty \d\a_1\ldots \d\a_r \vf(\a_1)\ldots \vf(\a_r) c(\a_1, \ldots,\a_r)$.
} 
we then have
\ba\label{ctcoefs}
c_\f(\L,\a_i)&=&c_\f^1(\a_i)\, \ln \L^2 A + c_\f^2(\a_i) \ ,\nonumber\\
c_R(\L,\a_i)&=&c_R^1(\a_i)\, \ln \L^2 A + c_R^2(\a_i) \ ,\nonumber\\
c_m(\L,\a_i)&=&c_m^1(\a_i)\, \L^2  \ln \L^2 A+ c_m^2(\a_i)\,  \L^2 + c_m^3(\a_i) \frac{1}{A} \  .
\ea
Note that counterterms involving explicitly the area $A$ like $c_\f^1,\ c_R^1,\ c_m^1$ and $c_m^3$ are non-local counterterms that should not occur in any standard QFT. However, we have already observed the similarity between the counterterms and the measure action. The latter actually corresponds to some well-defined values of $c_m^2$, $c_R^2$ and $c_m^3$, so we should certainly allow a non-vanishing counterterm coefficient $c_m^3$.

Of course, the counterterm action \eqref{Sct} cannot be expressed in terms of  geometric invariants written using only the metric $g$ and curvature $R$, contrary to the cosmological constant that is $\sim \int \d^2 x \sqrt{g}$. This is why people often consider that such counterterms should not be allowed. However, the whole quantization procedure is carried out with respect to some fixed background metric $g_0$ or $g_*$ and then one has $\wt\f=\frac{\k}{\sqrt{16\pi}}\left( 1 -\frac{\sqrt{g}}{\sqrt{g_*}}\right)$. In particular, already the Liouville action itself is expressed in terms of $\s$ which is defined by $g$ and $g_0$. Nevertheless, the Liouville action satisfies the cocycle condition\footnote{
The cocycle condition states that $S_L[g_0,\a+\b]=S_L[g_0,\b]+S_L[g_1,\a]$ if $g_1=e^{2\b} g_0$ and $g=e^{2\a} g_1$.
}
which is not the case for the measure action or for the counterterms.

The one-loop diagram with the counterterm vertices that follow from \eqref{Sct} then gives a contribution to be added to \eqref{Isum3} that is
\ba\label{Zct}
W^{(2)}[A]\Big\vert_{\rm ct}
&=&\frac{8\pi}{\k^2} \int\d x\left[-\frac{c_\f}{2} \wt K(t,x,x)- \left(  \frac{c_R}{2} R_* +\frac{c_m}{2} \right) \wh{\wt K}(t,x,x)\right] \nonumber \\
&&\hskip-2.cm =\ -\frac{1}{\k^2}\left[ c_\f \left( \frac{A\L^2}{\a}+\frac{7}{6}R_*A-4\pi\right)
+(c_R R_*A +c_m A) \left(G_0 +\ln \frac{A\L^2}{A_0\m^2} -\ln\a\right) \right]\ ,   \nonumber\\
\ea
where we used the small $t$ expansions \eqref{Ktilexplicit} and \eqref{Khattildexxexp} of $\wt K$ and $\wh{\wt K}$, as well as the relation \eqref{Adep} and the definition \eqref{alphafcts} for $G_0$.

Before actually computing the counterterm coefficients in the remainder of this section, let us  discuss what are the ``desired" values of the  $c_\f, c_R$ and $c_m$. In particular, the non-local terms $\sim A\L^2 (\ln A\L^2)^2$ and $ \sim (\ln A\L^2)^2$ are absent from $W^{(2)}[A]\big\vert_{\rm loops}$, as given in \eqref{Isum3} and, hence, should also be absent from \eqref{Zct}.
It is easy to see that this implies
\be\label{cm1cR1desired}
c_m^1 = c_R^1 = 0 \quad , \quad \text{(desired values)}\ .
\ee
Below, we will  indeed find that $c_m^1=c_R^1=0$, as well as $c_\f^1=0$. Anticipating \eqref{cm1cR1desired}, as well as\be\label{cphi1anti}
c_\f^1=0\ ,
\ee
eq. \eqref{Zct} becomes
\begin{multline}\label{W2ct}
W^{(2)}[A]\Big\vert_{\rm ct}
=-\frac{1}{\k^2} \Bigg\{
c_m^2
A\L^2 \ln A\L^2 +\left[c_m^3 
+ c_R^2R_*A\right] \ln A\L^2\\
+ \left[ \frac{c_\f^2}{\a}+c_m^2\left( G_0-\ln A_0\m^2-\ln \a \right)\right] A \L^2\Bigg\}
+ c_{\rm ct}(\a_i) +{\cal O}(1/\L^2) \ ,
\end{multline}
where $c_{\rm ct}(\a_i)$ is some $A$-independent finite function of the $\a_i$.
The terms in the second line are either $\sim A$ and can again be changed by adding an additional cosmological constant counterterm, or irrelevant area-independent constants.
If we add \eqref{W2ct} to \eqref{Isum3} we get
\begin{multline}\label{W2loopsW2ct}
W^{(2)}[A]\Big\vert_{\rm loops}+W^{(2)}[A]\Big\vert_{\rm ct}
=-\frac{1}{\k^2} \Bigg\{
\left[-\frac{1}{2\pi}H(\a_i)
+c_m^2\right] A\L^2 \ln A\L^2 \\
\hskip5.cm+\left[-\frac{1}{2\pi} F(\a_i)+c_m^3 
+ c_R^2R_*A\right] \ln A\L^2 \Bigg\}\\
+ A\L^2 \left[ \ldots\right] + c(\a_i)+c_{\rm ct}(\a_i) +{\cal O}(1/\L^2) \ .\hskip1.3cm
\end{multline}
We see that $c_\f^2$ only enters in the cosmological constant part and, hence, does not play any role. Cancellation of the $A\L^2 \ln A\L^2$ terms requires
\be\label{cphi1cm2desired}
c_m^2
=\frac{1}{2\pi}H(\a_i) \ .
\ee
Finally, the ``physical" coefficient of $\ln A\L^2$ should not depend on the choice of the regulator functions $\vf(\a_i)$, which requires
\be\label{alphaindep}
c_m^3 
+ c_R^2R_*A\equiv
c_m^3 
+8\pi(1-h) c_R^2
=\frac{1}{2\pi} F(\a_i) + {\rm const}\ ,
\ee
where ${\rm const}$ is a true, $\a_i$-independent constant. One could be tempted to equate separately the terms proportional to $R_*A\sim (1-h)$ and those not involving the curvature, resulting in ``universal", genus-independent counterterm coefficients. But since we compute on a surface of fixed genus $h$, this does not really make sense.
In any case, it is satisfying to remark  that, with the possible exception of $c_m^3$, the non-local counterterms $c_\f^1,\ c_m^1$ and $c_R^1$ are not required! After all these considerations, it is now time to actually  compute these coefficients. Remarkably, we will find that they satisfy all desired relations, in particular \eqref{cphi1cm2desired} and \eqref{alphaindep}, which are quite non-trivial.

\subsection{One-loop computation of the 2-point function and determination of the counterterm coefficients}

To actually determine the counterterms, we must do a one-loop computation of the 2-point Green's function and impose some convenient renormalization conditions to not only cancel the diverging parts but to also fix the finite contributions.

Note that in ordinary flat space quantum field theory one does not need to compute the full two-point Green's function $\Gd(u,v)$. Instead one rather  computes  the amputated or 1PI (which is the same at one loop) two-point function in momentum space. The corresponding contribution of the counterterms can then be read directly from the counterterm action. 
The analogue of such a momentum space computation is not available, in general, on a curved manifold,\footnote{
The analogue of momentum space is the mode decomposition with respect to the eigenfunctions $\p_r(x)$ of $\D$, replacing the plane waves $e^{ipx}$. An important property is momentum conservation that follows from $\int\d x e^{i(p_1+p_2+p_3)}\sim \dd(p_1+p_2+p_3)$. On a curved manifold one then needs the $C_{rst}=\int\d x \p_r \p_s \p_t$, etc. While on the round sphere, where the $\p_r$ are the spherical harmonics, the $C_{rst}$ are the well-known Clebch-Gordan coefficients, on the higher genus surfaces much less is known about the $C_{rst}$ and things are much more complicated.
} 
and we have to do the computation directly in real space. In this case it seems that the simplest way to correctly take into account all contributions is to compute $\Gd(u,v)$.

We will  compute the regularized\footnote{
Contrary to what we did at tree-level where we used $\wt G$ and $\wh{\wt K}$, resp $\wt G_{\vf,\L}$ to denote the Green's function and its regularized version, here we just write $\Gd$ since we will always deal with the regularized 2-point function.
}
2-point function $\Gd(u,v)$ which we define so that at tree-level it is just $\wh{\wt K}(t,u,v)$. At order $\k^{-2}$ it receives contributions at one loop and a contribution from the counterterms:
\be\label{G2contrib}
\Gd(u,v)=\wh{\wt K}(t,u,v) +  \Gd(u,v)\Big\vert_{\rm 1-loop}  
+ \Gd(u,v)\Big\vert_{\rm ct}  + {\cal O}(\k^{-4}) \ .
\ee 
A first renormalization condition we clearly want to impose on this full 2-point function is finiteness if $u\ne v\,$:
\be\label{RC1}
{\rm For}\ u\ne v\ \  : \quad \lim_{\L\to\infty} \Gd(u,v)\ \   \text{is finite.}
\ee
As we will see, this condition indeed allows us to determine all diverging parts of the counterterm coefficients, i.e. $c_\f^1,\ c_R^1,\ c_m^1$, as well as $c_m^2$.

It turns out to be surprisingly difficult to find a sensible renormalization condition to fix the finite parts of $\Gd$ and thus the finite parts of the counterterm coefficients. 
The trouble is that there is no analogue of a renormalization condition at some particular value of momentum. The only natural choice seems to be zero momentum, corresponding to the zero-mode of the two-point function. But $\int \d u\, \Gd(u,v)=0$ automatically, and this conditon is empty.
Instead, one might be tempted to try to impose some condition at $u=v$ like e.g.
\be
\label{RC2}
{\rm For}\ u=v\quad : \quad \lim_{\L\to\infty} \left[\Gd(u,u)- \wh{\wt K}(t,u,u)\right] 
\ \begin{matrix}?\\=\\{} \end{matrix}\  0 \ .
\ee
However, this doesn't make sense either. Indeed, the divergence of $\Gd(u,v)$  as $u\to v$ will turn out to be different from the one of  $\wh{\wt K}(t,u,v)$ : there are additional diverging and additional finite terms. As we will see, absence of the diverging terms would require a non-vanishing counterterm coefficient $c_\f^1$, which will be excluded  by requiring finiteness of the two-point function at $u\ne v$. Independently of this, the difference of the finite terms also makes it impossible to impose \eqref{RC2}.

Actually, this ``problem" was to be expected. The usual renormalization of the two-point function at some finite value of momentum, or equivalently at non-coinciding points, does not, of course yield a finite two-point function at coinciding points, nor does it imply that the loop and counterterm contributions to the two-point function vanish at coinciding points. This is the translation of the fact that to define composite operators like $(\wt \f(u))^2$ one needs an independent renormalization constant. Of course, in any ordinary flat-space quantum field theory the finite parts of the counterterm coefficients can be changed by changing the renormalization conditions -- this freedom being at the origin of the renormalization group. Hence, we should probably accept that we cannot (completely) fix the finite parts of our counterterm coefficients. In particular, this implies that, at the two-loop level, the finite coefficient of $\ln\frac{A}{A_0}$ in $W^{(2)}[A]-W^{(2)}[A_0]$ is a parameter that can be adjusted!

We will now present the computation of the regularized two-point function $\Gd(u,v)$ for $u\ne v$, and then  briefly give the result  for $u=v$. To keep this section readable, we have deferred many computational details to the appendix. One of the reasons we have to treat the cases $u\ne v$ and $u=v$ seperately is the following. In the case $u\ne v$ we can always choose $\L$ large enough so that   $\ell^2(u,v) \gg \frac{1}{\L^2}$. Making this assumption will simplify the evaluation of certain terms, as we will see shortly. 

\subsubsection{Counterterm contributions}

The regularized counterterm contribution to $\Gd(u,v)$ is easy to write down, since it only involves two regularized propagators $\wh{\wt K}(t_1,u,x)$ and $\wh{\wt K}(t_2,x,v)$ connected by the counterterm vertex as given by \eqref{Sct}:
\ba\label{G2ct}
\Gd(u,v)\Big\vert_{\rm ct} &=&\frac{1}{\k^2}\int \d x \Big[ -\frac{c_\f}{2} \wh{\wt K}(t_1,u,x) \wt K(t_2,x,v) 
 -\frac{c_\f}{2} {\wt K}(t_1,u,x) \wh{\wt K}(t_2,x,v)\nonumber\\
 && \hskip1.5cm  -(c_m + c_R\, R_*) \wh{\wt K}(t_1,u,x) \wh{\wt K}(t_2,x,v) \Big]\ .
\ea
We now assume that the counterterm coefficients have the general form \eqref{ctcoefs} without any further assumptions.
Note that, while $c_\f$ and $c_R$ involve at most an $\ln A\L^2$ divergence, the coefficient $c_m$ could have a $\L^2\ln A\L^2$ divergence. This means that, in order to compute the finite contributions, we should keep terms in $\wh{\wt K}(t_1,u,x) \wh{\wt K}(t_2,x,v)$ that are ${\cal O}(1/\L^2)$. (This is one of the reasons why all external propagators must also consistently be replaced by the regularized ones.)
A typical term we have to evaluate is the first one (where here we can replace $\wt K$ by $K$ since the $\frac{1}{A}$-piece yields a vanishing integral): $\int\d x\, \wh{\wt K}(t_1,u,x) K(t_2,x,v)$. As usual, at large $\L$, i.e. small $t_i$, the $e^{-(x-v)^2/4 t_2}$ in $K(t_2,x,v)$ forces $x$ to be close to $v$ (i.e. $(x-v)^2\sim t_2\sim \frac{1}{\L^2}$). 
Clearly, we will have to distinguish the cases $u=v$ and $u\ne v$. If $u=v$, we have to compute $\int \d x\, \wh{\wt K}(t_1,v,x) K(t_2,x,v)$ and we use the short distance expansion of $\wh{\wt K}$.
On the other hand, if  $u\ne v$ where we can suppose $\ell^2(u,v) \gg \frac{1}{\L^2}$, we can safely consider that $x$ is not close to $u$ and, hence, we can Taylor expand 
\be\label{KhatTaylor}
\wh{\wt K}(t_1,u,x)=\wh{\wt K}(t_1,u,v)+(x-v)^i \del^i_v \wh{\wt K}(t_1,u,v) +\frac{1}{2} (x-v)^i(x-v)^j \del^i_v\del^j_v \wh{\wt K}(t_1,u,v) + \ldots \ .
\ee
This can only be  a valid expansion as long as $\ell^2(u,x)\sim \ell^2(u,v) \gg \frac{1}{\L^2}$, so that $\wh{\wt K}(t_1,u,x)$ is a smooth function of $x$ in the vicinity of $v$. This is no longer the case if $u\to v$. In particular, if $\ell^2(u,x)\sim \ell^2(u,v)\sim \frac{1}{\L^2}$, all terms in the expansion \eqref{KhatTaylor} are similarly large.

Then, let us first evaluate \eqref{G2ct} for $u=v$. It is straightforward to obtain
\ba\label{G2ctuu}
\Gd(u,u)\Big\vert_{\rm ct} &=& - \frac{c_\f}{\k^2} \wh{\wt K}(t,u,u) +\frac{c_\f}{4\pi\k^2} \ln\frac{\a_1+\a_2}{\a}\nonumber \\ 
&& -\frac{c_m+c_R R}{\k^2} \int \d x\, \wh{\wt K}(t_1,u,x) \wh{\wt K}(t_2,x,u)+{\cal O}(1/\L^2) \ .
\ea
Note that the integral is a finite smooth function of $u$ even in the limit $t_i\to 0$. 
However, as mentioned above, one needs to keep the $t_i$ finite, since the subleading terms, multiplied with the divergent pieces from $c_m$, can lead to  finite contributions.

Next, for  $\ell^2(u,v) \gg \frac{1}{\L^2}$ we get similarly
\ba\label{G2ctuv}
\hskip-1.cm u\ne v : \quad \Gd(u,v)\Big\vert_{\rm ct} &=& - \frac{c_\f}{\k^2} \wh{\wt K}(t,u,v)\nonumber\\
&&  -\frac{c_m+c_R R}{\k^2} \int \d x\,   \wh{\wt K}(t_1,u,x) \wh{\wt K}(t_2,x,v)+{\cal O}(1/\L^2) \ .\ \ \
\ea
Of course, one is not allowed to simply take the $u\to v$ limit of this expression to get $\Gd(u,u)$ and, indeed, \eqref{G2ctuu} differs by a term $\sim\ln\frac{\a_1+\a_2}{\a}$ from the naive limit. Here, for $u\ne v$ the function $\wh{\wt K}(t,u,v)$ is non-singular for $t\to 0$. More precisely, it equals $\wt G(u,v)$ plus terms that are either exponentially small or of order $1/\L^2$. Since $c_\f$ is at most logarithmically divergent, we may replace $c_\f \wh{\wt K}(t,u,v)$ by $c_\f \wt G(u,v)$ up to terms that vanish as $\L\to\infty$. Thus
\ba\label{G2ctuv2}
\hskip-1.cm  u\ne v : \quad \Gd(u,v)\Big\vert_{\rm ct} &=& - \frac{c_\f}{\k^2} \wt G(u,v)\nonumber\\
&& \hskip-1.cm -\frac{c_m+c_R R}{\k^2} \int \d x\,   \wh{\wt K}(t_1,u,x) \wh{\wt K}(t_2,x,v)+{\cal O}(\ln\L^2/\L^2) \ .
\ea
We see that the finite  ${\cal O}(\L^0)$ parts of $c_m$ and $c_R$, i.e. $c_R^2$ and $c_m^3$ only give finite contributions to $\Gd\Big\vert_{\rm ct}$. Hence, to determine them, we need to obtain the finite ${\cal O}(\L^0)$ terms in $\Gd\Big\vert_{\rm 1-loop}$.


\def\diaguOv{\begin{picture}(26,10)(0,0)
\put(0,3.5){\line(1,0){5}}\put(10,3.5){\circle{10}}\put(15,3.5){\line(1,0){5}}
\end{picture}}
\def\diaguTOv{\begin{picture}(26,20)(0,0)
\put(0,3.5){\line(1,0){20}}\put(10,15.5){\circle{10}}\put(10,3.5){\line(0,0){7}}
\end{picture}}
\def\diaguTOvsmall{\begin{picture}(10,10)(0,0)
\put(0,1.5){\line(1,0){10}}\put(5,6){\circle{4}}\put(5,1.5){\line(0,0){2.5}}
\end{picture}}
\def\tadpole{\begin{picture}(26,20)(0,0)
\put(10,15.5){\circle{10}}\put(10,5.5){\line(0,0){5}}
\end{picture}}
\def\diaguIOv{\begin{picture}(26,20)(0,0)
\put(0,3.5){\line(1,0){20}}\put(10,8.9){\circle{10}}
\end{picture}}
\def\diaguIOvsmall{\begin{picture}(10,10)(0,0)
\put(0,1.5){\line(1,0){10}}\put(5,5){\circle{6}}
\end{picture}}
%
\subsubsection{Total one-loop contribution to  $\Gd(u,v)$ for $u\ne v$}

We will now determine the one-loop contributions to $\Gd(u,v)$ for $u\ne v$. As already emphasized, we may assume that $\ell^2(u,v)\gg \frac{1}{\L^2}$.
There are three one-loop diagrams contributing to $\Gd(u,v)$. They are $\diaguOv$, $\diaguTOv$ and $\diaguIOv$ (including the regularized external propagators). These diagrams contribute at the same order $\sim\frac{1}{\k^2}$ as the tree contribution from the measure vertex  and the one from the counterterms. The counterterm contribution has been determined above in \eqref{G2ctuv2}. The computation of the three two-point one-loop diagrams  is quite lengthy, mainly due to the non-symmetric nature of the cubic and quartic vertices. This implies that there are many different contributions to each diagram. We have deferred the details of the computation to the appendix \ref{appG2unev} where the result for the  three one-loop diagrams is given in
\eqref{u-O-v2}, \eqref{GuTOv2} and  \eqref{GuIOv2} and for  the measure contribution in \eqref{G2measure}.

We now add these contributions  \eqref{u-O-v2}, \eqref{GuTOv2},  \eqref{GuIOv2} and \eqref{G2measure}, as well as the counterterm contribution \eqref{G2ctuv2}.
As before, since these expressions have to be multiplied with $\int \d\a_i \vf(\a_i)$ for every $\a_i$, we may symmetrize all expressions in $\a_i$.
We get
\ba
&&\hskip-0.5cm\left(\Gd_{u-\hskip-0.8mm{\rm O}\hskip-0.8mm-v}+
\Gd_{u\, \diaguTOvsmall \,v}+
\Gd_{u\, \diaguIOvsmall \,v}+
\Gd_{\rm measure}+
\Gd\Big\vert_{\rm ct}\right)(u,v) 
\nonumber\\
&&\hskip-0.5cm=\frac{1}{2\k^2}\Bigg\{
\frac{1}{\pi}\wt G(u,v) \left[\frac{3}{2}\ln\frac{\a_2\a_3}{(\a_2+\a_3)^2 }-1-\frac{2\a_2\a_3}{(\a_2+\a_3)^2}  - 2\pi c_\f^1 \ln A\L^2 - 2\pi c_\f^2\right]
\nonumber\\
&&+2\wt G(u,v)\left[ \wt G_\zeta(u)+\wt G_\zeta(v)-\frac{2}{A}\int\d x\, \wt G_\zeta(x)\right] 
-\frac{2}{A}\int\d y \big[\wt G(u,y)+ \wt G(y,v)\big)] \wt G_\zeta(y)
\nonumber\\
&&+\frac{1}{\pi}\int \d y \wh{\wt K}(t_1,u,y)\wh{\wt K}(t_4,y,v) 
\Bigg[\frac{2\L^2}{\a_2+\a_3} -\frac{5\L^2}{2\a_2}-2\pi c_m^1 \L^2\ln A\L^2 -2\pi c_m^2 \L^2
\nonumber\\
&&\hskip5.mm -\frac{10\pi}{A} -\frac{2\pi c_m^3}{A}
+R_*\left( \frac{3}{2}\ln\frac{\a_2\a_3}{(\a_2+\a_3)^2}-\frac{19}{12} -\frac{2\a_2\a_3}{(\a_2+\a_3)^2} -2\pi c_R^1 \ln A\L^2 -2\pi c_R^2 \right) 
\nonumber\\
&&\hskip5.mm +2\pi R_* \left(\wt G_\zeta(y) -\frac{1}{A}\int\d x\, \wt G_\zeta(x) \right)\Bigg] 
\nonumber
\ + 
\ea
\ba
&&+2(\wt G(u,v))^2
-\frac{2}{A}\int\d y (\wt G(u,y))^2 
-\frac{2}{A}\int\d y (\wt G(y,v))^2 
+\frac{2}{A^2}\int\d x \d y (\wt G(x,y))^2 
\nonumber\\
&&+2R_*\int\d y\, \left[ (\wt G(u,y))^2\,\wt G(y,v) +\wt G(u,y) (\wt G(y,v))^2\right] 
\nonumber\\
&&-\frac{8R_*}{A} \int\d x\d y\, \wt G(u,x)\wt G(x,y) \wt G(y,v)
-\frac{2R_*}{A}\int\d x\d y\,  (\wt G(x,y))^2 \left[ \wt G(u,x) +\wt G(y,v)\right] 
\nonumber\\
&&+2R_*\,  \wt G(u,v) \int\d x\, \left(\wt G(u,x)+\wt G(v,x)\right) \wt G_\zeta(x) 
\nonumber\\
&&-\frac{2R_*}{A} \int\d x\, \left(\wt G(u,x)+\wt G(v,x)\right)\int\d y\, \wt G(x,y) \wt G_\zeta(y)\ +
\nonumber\\
&&+2R_*^2 \int\d x\d y\, \left[ \wt G(u,x) \,(\wt G(x,y))^2\, \wt G(y,v)+\wt G(u,x)\wt G(x,v)\wt G(x,y) \wt G_\zeta(y) \right]
 \Bigg\}\nonumber\\
&&+{\cal O}(1/\L^2) \ . \label{totaloneloop}
\ea
Finiteness requires
\be\label{c1values}
c_\f^1=c_m^1=c_R^1=0\quad , \quad
c_m^2=\frac{1}{2\pi} \left( \frac{2}{\a_2+\a_3} -\frac{5}{2\a_2}  \right) \ .
\ee
These are exactly the ``desired values"  \eqref{cm1cR1desired},  \eqref{cphi1anti} and
\eqref{cphi1cm2desired}. In particular, the value of $c_m^2$ is exactly what is needed to cancel the divergent two-loop contributions in $W[A]=\ln Z[A]\,$! As explained above, the value \eqref{c1values} for $c_m^2$  really means that $c_m^2[\vf]=\int_0^\infty \d\a_2 \d\a_3 \vf(\a_2)\vf(\a_3)\frac{1}{2\pi} \left( \frac{2}{\a_2+\a_3} -\frac{5}{2\a_2}  \right)$.

Next, since there are no more divergent coefficients, we may now safely replace the 
$\wh{\wt K}(t_1,u,y)\wh{\wt K}(t_4,y,v) $ by the regulator independent $\wt G(u,y)\wt G(y,v)$.
We moreover require that the Green's function $\Gd(u,v)$ should not depend at all on the regulator functions $\vf(\a)$, i.e. that  \eqref{totaloneloop} should not depend on the $\a_i$. This yields
\ba\label{cf2value}
c_\f^2&=&\frac{1}{2\pi}\left[\frac{3}{2}\ln\frac{\a_2\a_3}{(\a_2+\a_3)^2 }-1-\frac{2\a_2\a_3}{(\a_2+\a_3)^2} \right] +\wh c_\f\ , \\
\label{cR2value}
c_R^2&=&\frac{1}{2\pi}\left[\frac{3}{2}\ln\frac{\a_2\a_3}{(\a_2+\a_3)^2 }-\frac{19}{12}-\frac{2\a_2\a_3}{(\a_2+\a_3)^2} \right] +\frac{\wh c_R}{2\pi} \ ,\\
\label{cm3value}
c_m^3&=& \wh c_m\ ,
\ea
where $\wh c_\f$, $\wh c_R$ and $\wh c_m$ are true ($\a_i$-independent, $\L$-independent) constants. With these values of the counterterm coefficients, eq. \eqref{totaloneloop} reduces to
\ba
&&\hskip-0.5cm\left(\Gd_{u-\hskip-0.8mm{\rm O}\hskip-0.8mm-v}+
\Gd_{u\, \diaguTOvsmall \,v}+
\Gd_{u\, \diaguIOvsmall \,v}+
\Gd_{\rm measure}+
\Gd\Big\vert_{\rm ct}\right)(u,v) 
\nonumber\\
&&\hskip-0.5cm=\frac{1}{2\k^2}\Bigg\{
-2\,\wh c_\f \,\wt G(u,v)\nonumber\\
&&+2\wt G(u,v)\left[ \wt G_\zeta(u)+\wt G_\zeta(v)-\frac{2}{A}\int\d x\, \wt G_\zeta(x)\right] 
-\frac{2}{A}\int\d y \big[\wt G(u,y)+ \wt G(y,v)\big)] \wt G_\zeta(y)
\nonumber\\
&&+2\int \d y\, \wt G(u,y) \wt G(y,v)
\Bigg[ -\frac{\wh c_m+5}{A}
+ R_* \left(-  \frac{\wh c_R}{2\pi}+\wt G_\zeta(y) -\frac{1}{A}\int\d x\, \wt G_\zeta(x) \right)\Bigg]
\nonumber\\
&&+2(\wt G(u,v))^2
-\frac{2}{A}\int\d y (\wt G(u,y))^2 
-\frac{2}{A}\int\d y (\wt G(y,v))^2 
+\frac{2}{A^2}\int\d x \d y (\wt G(x,y))^2 
\nonumber\\
&&+2R_*\int\d y\, \left[ (\wt G(u,y))^2\,\wt G(y,v) +\wt G(u,y) (\wt G(y,v))^2\right] 
\nonumber \ +
\ea
\ba
&&-\frac{4R_*}{A} \int\d x\d y\, \wt G(u,x)\wt G(x,y) \wt G(y,v)
-\frac{2R_*}{A}\int\d x\d y\,  (\wt G(x,y))^2 \left[ \wt G(u,x) +\wt G(y,v)\right] 
\nonumber\\
&&+2R_*\,  \wt G(u,v) \int\d x\, \left(\wt G(u,x)+\wt G(v,x)\right) \wt G_\zeta(x) 
\nonumber\\
&&-\frac{2R_*}{A} \int\d x\, \left(\wt G(u,x)+\wt G(v,x)\right)\int\d y\, \wt G(x,y) \wt G_\zeta(y)
\nonumber\\
&&+2R_*^2 \int\d x\d y\, \left[ \wt G(u,x) \,(\wt G(x,y))^2\, \wt G(y,v)+\wt G(u,x)\wt G(x,v)\wt G(x,y) \wt G_\zeta(y) \right]
 \Bigg\}\nonumber\\
&&+{\cal O}(1/\L^2) \label{totaloneloop2} \ ,
\ea
which now is finite and completely regulator independent. As a consistency check, we note that this expression correctly vanishes if integrated over $\d u$ or over $\d v$.

There seems to be no obvious way to fix the remaining constants $\wh c_\f$, $\wh c_R$ and $\wh c_m$ without imposing some definite value for $\Gd(u,v)$ at some fixed $u,v$.  As already discussed, imposing a condition at  $u=v$ does not help since in this case new divergences appear that have to be renormalized independently.
In any case, our  expression \eqref{totaloneloop2} is only valid for $u\ne v$.

\subsubsection{Total one-loop contribution to $\Gd(u,u)$}

To explicitly see which are the new divergences and finite terms that appear for $u=v$, as well as for possible future reference, we now quote the result for the total one-loop plus counterterm contributions to the two-point function at coinciding points $u=v$. This case  involves some interesting technical difficulties but since we do not use this any further we do not spell out the computation and only give the result:
\begin{multline}\label{totaloneloopuu}
\left(\Gd_{u-\hskip-0.8mm{\rm O}\hskip-0.8mm-u}+
\Gd_{u\, \diaguTOvsmall \,u}+
\Gd_{u\, \diaguIOvsmall \,u}+
\Gd_{\rm measure}+
\Gd\Big\vert_{\rm ct}\right)(u,u) 
\\
\hskip-0.cm=\frac{1}{2\k^2}\Bigg\{
2 \wh{\wt K}(t_i,u,u)\wh{\wt K}(t_j,u,u)  + \left[\hat c_1(\a_i)-2 c_\f\right] \wh{\wt K}(t_i,u,u) +\left[\hat c_2(\a_i)+\frac{c_\f}{2\pi} \ln\frac{\a_i+\a_j}{\a_i}\right]\\
+4\wh{\wt K}(t_1+t_4,u,u) \Big[ \wt G_\zeta(u) -\frac{1}{A}\int\d x\, \wt G_\zeta(x)+R_*\, \int\d x\, \wt G(u,x) \wt G_\zeta(x)\Big]\\
\hskip1.5cm+\frac{1}{\pi}\int \d y \wh{\wt K}(t_1,u,y)\wh{\wt K}(t_4,y,u) 
\Big[\frac{2\L^2}{\a_2+\a_3} - \frac{5\L^2}{2\a_2} -2\pi c_m \hskip5.cm\\
\hskip3.cm+R_*\Big(\frac{3}{2}\ln\frac{\a_2\a_3}{(\a_2+\a_3)^2}
-\frac{2\a_2\a_3}{(\a_2+\a_3)^2}-\frac{19}{12}-2\pi c_R\Big)  \Big]
\\  
-\frac{14}{A}\int\d y\, (\wt G(u,y))^2 
-\frac{4}{A}\int\d y\, \wt G(u,y) \wt G_\zeta(y) 
+\frac{2}{A^2}\int\d x \d y\, (\wt G(x,y))^2  \ + \nonumber
\end{multline}
\begin{multline}
\hskip1.cm +4R_*\int\d y\, (\wt G(u,y))^3 + 2 R_* \int\d y\, (\wt G(u,y))^2\wt G_\zeta(y) \\
-\frac{4R_*}{A} \int\d x\d y\, \wt G(u,x)\wt G(x,y) \wt G(y,u)
-\frac{4R_*}{A}\int\d x\d y\, \wt G(u,x)  (\wt G(x,y))^2 \\
-\frac{4R_*}{A}\int\d x\d y\, \wt G(u,x)  \wt G(x,y) \wt G_\zeta(y)
-\frac{2R_*}{A}\int\d x\d y\, (\wt G(u,x) )^2  \wt G_\zeta(y)\\
\hskip-4.cm+2R_*^2 \int\d x\d y\, \wt G(u,x) \,(\wt G(x,y))^2\, \wt G(y,u)\\
+2R_*^2 \int\d x\d y\, (\wt G(u,x))^2 \,\wt G(x,y)\, \wt G_\zeta(y)\, +{\cal O}(\L^{-2}) 
\Bigg\} \ .\hskip2.5cm
\end{multline}
Here, $\hat c_1(\a_i)$ and $\hat c_2(\a_i)$ are finite coefficients that we did not determine.\footnote{The first coefficient $\hat c_1(\a_i)$ multiplying $\wh{\wt K}(t_i,u,u) $ can be absorbed into the (finite part of the) counterterm coefficient $c_\f$. It is possible (but  by no means obvious from our computation) that this same choice also cancels the second undetermined coefficient $\hat c_2(\a_i)$. }

As discussed above, one should not expect finiteness of $\Gd(u,u)$ or that it equals $\wh{\wt K}(t,u,u)$ as $\L\to\infty$. Indeed, the first term in \eqref{totaloneloopuu}, i.e.
$2\wh{\wt K}(t_i,u,u) \wh{\wt K}(t_j,u,u)$ is divergent as $\L\to\infty$. The same is true for the terms of the second line  of the right hand side. Cancelling these divergences would require a non-vanishing counterterm coefficient $c_\f^1$, which was excluded above by demanding finiteness of the two-point function at $u\ne v$. Independently of the value of $c_\f^1$,  one clearly cannot require that $\Gd(u,u)$ equals $\wh{\wt K}(t,u,u)$ in the $\L\to\infty$ limit.

\section{Final result and discussion}


We have done a careful evaluation of the two-dimensional quantum gravity partition function at fixed area up to two loops and including all contributions from the non-trivial measure and counterterms that contribute at the same order. We worked in the K\"ahler formalism that is well adapted at fixed area and used the general spectral cutoff regularization which works well on curved manifolds for multi-loop diagrams. The contributions of the counterterms turned out to be crucial. They were consistently determined from the finiteness of the one-loop two-point Green's function of the K\"ahler field.

The final result for the logarithm of the partition function is obtained from inserting the values of the counterterm coefficients we have determined in \eqref{c1values}-\eqref{cm3value} into \eqref{W2loopsW2ct}:
\begin{multline}\label{W2loopsW2ct-2}
W^{(2)}[A]\Big\vert_{\rm tot}=W^{(2)}[A]\Big\vert_{\rm loops}+W^{(2)}[A]\Big\vert_{\rm ct}
=-\frac{1}{\k^2} 
\Big[\wh c_m + 1 +4(1-h) \wh c_R\Big] \ln A\L^2 \\
+ A\L^2 \left[ \ldots\right] + c(\a_i)+c_{\rm ct}(\a_i) +{\cal O}(1/\L^2) \ .\hskip1.3cm
\end{multline}
We see that not only the terms $\sim A\L^2 \ln A\L^2$ have canceled, moreover the coefficient of $\ln A\L^2$ now is independent of the $\a_i$, i.e. independent of the regulator functions $\vf(\a)\,$! It only depends on the two finite renormalization constants  $\wh c_m$ and $\wh c_R$.
As already repeatedly emphasized, we can also adjust the cosmological constant counterterm to cancel any divergence in $W$ that is proportional to the area $A$. Subtracting  the same expression evaluated at area $A_0$, we finally get
\begin{multline}\label{W2loopsW2ct-3}
\ln \frac{Z[A]}{Z[A_0]}\Bigg\vert_{\k^{-2}}=W^{(2)}[A]\Big\vert_{\rm tot}-W^{(2)}[A_0]\Big\vert_{\rm tot}\\
=-\frac{1}{\k^2} 
\Big[\wh c_m + 1 +4(1-h) \wh c_R\Big] \ln \frac{A}{A_0} 
+ (A-A_0)\L^2 \left[ \ldots\right] +{\cal O}(1/\L^2) \ .\hskip1.3cm
\end{multline}
Equivalently, this shows that the area dependence of the partition function is
\be\label{finalconcl}
\ln Z[A]\Big\vert_{\rm two-loop + measure + ct}\sim e^{-\m_c^2 A} \, A^{\g_{\rm str}^{(2)} -3}\ ,
\ee
with 
\be\label{gammastr2loop}
\g_{\rm str}^{(2)} =\frac{2}{\k^2} \left[-2\,\wh c_R\,(1-h) -\frac{\wh c_m+1}{2}\right] \ .
\ee

We see that our careful, first-principles computation of the 2D quantum gravity partition function has established that, up to two loops, the partition function has indeed the expected form \eqref{finalconcl}. However, we have also found a maybe unexpected dependence on two finite renormalization constants.
By which principle should these counterterm coefficients be fixed? We observe that our result is compatible with the KPZ scaling, since we get  agreement with the (two-loop prediction of the) KPZ formula if we choose
\be\label{KPZvalues}
\wh c_m\Big\vert_{\rm KPZ}=-1\quad , \quad \wh c_R\Big\vert_{\rm KPZ}=-\frac{1}{2} \ .
\ee

In the absence of any principle to fix these constants, the area-dependence of the partition function $Z[A]$ appears to involve an arbitrary power of $A$. One more principle we may invoke is locality. Locality of the counterterms implies that all coefficients $c_\f^1,\ c_m^1,\ c_R^1$ which multiply a non-local $\ln A\L^2$ should vanish. We have indeed found this. But it would also imply that $c_m^3=\wh c_m$ which multiplies a non-local $\frac{1}{A}$ should equally vanish. On the other hand, such a $\frac{1}{A}$-term was already present in the  regularized measure action  due to the absence of a zero-mode. Indeed, from \eqref{Smeasure2} or \eqref{measurediag} we read that
\be\label{Smeasurereg}
S_{\rm measure}=-\frac{4\pi}{\k^2} \int \d^2 x\, \sqrt{g_*}\, \left[ \frac{1}{4\pi t} +\frac{7}{24 \pi} R_* -\frac{1}{A} +\ldots\right]\wt\f^2(x) \ .
\ee
Thus, a $\frac{1}{A}$-counterterm should certainly also be allowed. However, we may require  that the non-local   $\frac{1}{A}$-counterterm should actually cancel the non-local $\frac{1}{A}$-term in $S_{\rm measure}$. Comparing \eqref{Smeasurereg} with \eqref{Sct}, \eqref{ctcoefs} we see that this cancellation requires the following condition\be\label{cm3value-2}
\text{absence of non-local $\frac{1}{A}$-terms in $S_{\rm measure}+S_{\rm ct}$}
\quad \Rightarrow \quad
c_m^3\equiv \wh c_m = -1 \ .
\ee 
This is precisely the KPZ-value \eqref{KPZvalues}\,!

Which other requirement can we impose to also fix $\wh c_R\,$?
An important issue in quantum gravity is background independence: physical results should not depend on the choice of background metric $g_0$. This is certainly true for our final result \eqref{finalconcl}, \eqref{gammastr2loop}, for all values of $\wh c_R$ (and of $\wh c_m$). Of course, we cannot expect the counterterms to be background independent. Already to write the Liouville action we must choose a background metric, although the Liouville action satisfies the important cocycle condition, i.e. if $g=e^{2\a} g_1$ and $g_1=e^{2\b} g_0$ one has $S_L[g_0,\a+\b]=S_L[g_0,\b]+S_L[g_1,\a]$, which translates that it originated from a background independent definition of quantum gravity.  On the other hand, to spell out the functional integral measure in terms of the modes of $\D_0$ we must again introduce the reference metric $g_0$ and then neither the non-trivial measure action, nor the counterterm action can be explicitly
background independent or satisfy some analogous cocycle identity.

To summarize, there does not seem to be an obvious criterion why to choose  the KPZ-value $\wh c_R=-\frac{1}{2}$ rather than any other value. We are then led to consider that there could be different choices of $\wh c_R$ leading to consistent quantization schemes of this two-dimensional gravity. One could imagine that at least some of these new schemes could be consistent quantum gravities for all matter central charges, thus allowing to go beyond the $c=1$ barrier. However, the latter is invisible in a perturbative expansion in $\k^{-2}$, and it is premature to draw any conclusion in this direction.

Let us discuss what we can expect beyond the two-loop computation of this paper.
Standard power counting shows that {\it any}  loop-diagram  has a superficial degree of divergence equal to 2. Indeed, as is well-known, it is a particularity of scalar theories in two dimensions that the superficial degree of divergence of a diagram does not depend on the number of external lines. From what we have seen it is clear that this quadratic divergence $A \L^2$ can be accompanied by powers of $\ln A \L^2$. In particular, we expect that the leading divergence of an $L$-loop  diagram is $\sim A\L^2 (\ln A\L^2)^{L-1}$. Indeed, consider a diagram with $I$ propagators and $V$ vertices. Each internal line contributes a regularized propagator $\wh{\wt K}\sim 4\pi \wt G_\zeta-\ln t_i \sim\ln A\L^2$. Each vertex contributes a Laplacian which, when acting on $\wh{\wt K}$ converts the leading $\ln \L^2 A$ into a $\L^2$. Each vertex also contributes an integration over the manifold. For the leading singularity this contributes a factor $A \prod_{i=1}^{V-1}t_i\sim A (\frac{1}{\L^2})^{V-1}$. Thus, we get for the leading singularity of an $L$-loop vacuum diagram an overall factor of 
\be\label{Lloopdiv}
(\ln A\L^2)^{I-V} (\L^2)^V  A  (\frac{1}{\L^2})^{V-1}=(\ln A\L^2)^{L-1} A \L^2\ .
\ee

For example, at three loops, in addition to the cubic and quartic vertices, we also need to take into account quintic and sextic vertices, both involving one Laplace operator. The sextic vertex e.g. gives rise to a contribution $\sim \int\d x \wt K \wh{\wt K} \wh{\wt K}$ resulting in a leading divergence $\sim A\L^2 \left( \ln A \L^2\right)^2$.  Another 3-loop contribution comes from two quartic vertices joined by four propagators. While naively the leading divergence of this diagram may be thought to be $\sim A\L^2 \left( \ln A \L^2\right)^3$, closer inspection shows that it is $\sim A\L^2 \left( \ln A \L^2\right)^2$ (cf the discussion after \eqref{J11}), in agreement with the above general discussion.  At this same order, one also has to include a cubic and quartic vertex from the measure. The latter e.g. is $\sim \L^2$ and contributes,  through a two-loop ``figure-eight" diagram, again a leading divergence $\sim A\L^2 \left( \ln A \L^2\right)^2$. Further contributions come from two-loop diagrams including the one-loop counterterms determined above. Unless all these divergences somehow ``miraculously" cancel, their presence almost certainly will require the introduction of new counterterms, e.g.~cubic and quartic counterterm vertices with  coefficients $\sim\L^2$. Just as the corresponding measure vertices, such  counterterms then give  two-loop contributions that could cancel the $\sim A\L^2 \left( \ln A \L^2\right)^2$ divergences. We have carried out some preliminary investigations about these leading divergences at this three-loop level \cite{LLAB} and we do not see any ``miraculously" cancellation. Thus, new cubic and quartic counterterms seem indeed to be required.\footnote{ 
Of course, one could also cancel these leading divergences through one-loop diagrams including the counterterms  \eqref{Sct} by adding non-local higher-order contributions  $\sim\frac{1}{\k^4} \L^2 \ln A\L^2$ to their coefficients. However, as argued above, this is against the principles of local quantum field theory and such coefficients certainly do not correspond to the cancellation of some divergence in the two-loop two-point function.
}
Again, the diverging parts of these counterterm coefficients could then be determined from requiring finiteness of the one-loop three-point and four-point functions, while their finite parts possibly remain undetermined.  Of course, there will also be new $A\L^2 \ln A \L^2$ divergences from the three-loop diagrams, but they can be cancelled simply by additional order $\k^{-4}$-contributions to the quadratic counterterms.

The same structure will continue, at $L$ loops, where the leading divergence is given by \eqref{Lloopdiv}. It is thus reasonable to expect that at every order in perturbation theory all unwanted, non-local divergences can be removed by appropriate local counterterms. This will also lead to the introduction of new finite renormalization constants  and we clearly expect that the finite coefficient of $\ln \frac{A}{A_0}$ in  $\ln \frac{Z[A]}{Z[A_0]}$ depends on these finite constants. While eventually this might open the way to new quantization schemes that could allow to circumvent the $c=1$ barrier, obviously the r\^ole of these constants needs to be better understood. 


\begin{appendix}
%
\section{Appendix}

\subsection{Integrals}\label{integrals}

Here we list various integrals of the form
\be
I[f]=\frac{1}{\pi}\int \d^2 \wt z\ e^{-\wt z^2} f(\wt z^2) = \int_0^\infty \d \xi\ e^{-\xi} f(\xi) \ ,
\ee
where $\d^2 \wt z$ is the flat measure: any non-trivial expansion of $\sqrt{g}$ is included in $f$. Then
\ba
&& I[\xi^n]=n! \quad , \quad 
 I[\xi^n \ln \xi ]=c_n - (n!)\, \g\quad (c_0=0,\, c_1=1,\, c_2=3,\, c_3=11)
\nonumber\\
&& I[(\ln\xi)^2]=\g^2+\frac{\pi^2}{6}
\quad , \quad
I[\xi (\ln\xi)^2]=\g^2-2\g+\frac{\pi^2}{6}
\nonumber\\
&& I[E_1(\xi/a)]=\ln(1+a)
\quad , \quad I[\xi E_1(\xi/a)]=-\frac{a}{1+a}+\ln(1+a)
\nonumber\\
&& I[\xi^2 E_1(\xi/a)]=-\frac{a(2+3a)}{(1+a)^2}+2\ln(1+a)
\nonumber
\ea
\ba
&&I[\xi^3 E_1(\xi/a)]=-\frac{a(11 a^2+15 a+6)}{(1+a)^3}+6\ln(1+a)
\nonumber\\
&&I[E_2(\xi/a)]=1-\frac{1}{a}\ln(1+a)
\quad , \quad
 I[\xi E_2(\xi/a)]=1+\frac{1}{1+a}-\frac{2}{a} \ln(1+a)
\nonumber\\
&& I[\ln\xi\ E_1(\xi/a)]=-\frac{\pi^2}{6}-\g\ln(1+a)+Li_2(\frac{1}{1+a})
\ea
Of course, insertion of any odd number of components of $\wt z$ into any of these integrals gives a vanishing result, while insertions of an even number can be replaced according to the usual rules
$\wt z^i\wt z^j \to \frac{1}{2}\, \wt z^2\,  \delta^{ij}$ and 
$\wt z^i\wt z^j \wt z^k \wt z^l \to \frac{1}{8}\, (\wt z^2)^2\, \left( \delta^{ij}\delta^{kl}+\delta^{ik}\delta^{jl}+\delta^{il}\delta^{jk}\right)$.

\subsection{One-loop contributions to the two-point function $\Gd(u,v)$}\label{appG2unev}

In this appendix we give some details of the computation of the one-loop contributions to the two-point function  $\Gd(u,v)$ for non-coinciding points $u\ne v$.

\subsubsection{One-loop contribution from $u-\hskip-1.3mm{\rm O}\hskip-1.3mm-v$}

Recall that $\Gd(u,v)$ always includes the two external propagators that, by consistency, are also regularized, i.e. replaced by $\wh{\wt K}$. Then, for the diagram  $\diaguOv$,
due to the many ways the derivatives in the cubic vertex can act, one gets many different contributions. They yield
\begin{multline}\label{u-O-v}
\Gd_{u-\hskip-0.8mm{\rm O}\hskip-0.8mm-v}(u,v)
=\frac{1}{2\k^2}\int\d x\, \d y\Big[
2\wt K \wh{\wt K} \wh{\wt K}\wt K +4 \wt K  \wt K  \wh{\wt K}\wh{\wt K} + 4  \wh{\wt K} \wh{\wt K} \wt K \wt K + 4 \wh{\wt K} \wt K\wt K  \wh{\wt K} \\
+ 4  \wh{\wt K}  \wh{\wt K} (-\frac{\d}{\d t} \wt K )  \wh{\wt K} + 2 R_* \wt K  \wh{\wt K} \wh{\wt K} \wh{\wt K} + 2 R_*  \wh{\wt K} \wh{\wt K} \wh{\wt K}\wt K + 8 R_*  \wh{\wt K} \wt K  \wh{\wt K} \wh{\wt K} + 2 R_*^2  \wh{\wt K} \wh{\wt K} \wh{\wt K} \wh{\wt K} \Big] ,
\end{multline}
where 
$\wt K \wh{\wt K} \wh{\wt K}\wt K$ stands for $\wt K(t_1,u,x) \wh{\wt K}(t_2,x,y) \wh{\wt K}(t_3,x,y) \wt K(t_4,y,v)$ etc.

We now evaluate this for $u\ne v$. More precisely, since we work at finite cutoff, we do not want $\ell^2(u,v)$ to be as small as $\frac{1}{\L^2}$ and require $\ell^2(u,v)\, \L^2\gg 1$. Then $K(t=\a/\L^2,u,v)\sim \frac{\L^2}{4\pi\a} e^{-\ell^2 \L^2/(4\a)}$ is exponentially small and can always be dropped. Also $\wh{\wt K}(t,u,v)=\wt G(u,v) +\text{exponentially small}+{\cal O}(\frac{1}{\L^2})$, cf \eqref{KhatGint} or \eqref{Khattilsmallt}. Furthermore, in $\int\d y \wh{\wt K}(t_1,u,y)\wh{\wt K}(t_2,y,v)$ or in $\int\d y \wh{\wt K}(t_1,u,y)\wh{\wt K}(t_2,y,u)$
we may replace the $\wh{\wt K}$ by $\wt G$ since these integrals have finite limits as $\L\to\infty$. (The logarithmic short-distance singularity $(\ln \m^2(y-u)^2)^n$ is integrable for any integer $n$.)

Denoting by $+\ldots$ terms that are either ${\cal O}(1/\L^2)$ or exponentially small as just explained, we find
\begin{multline}\label{kkhatkhatk}
\int\d x\, \d y\, \wt K \wh{\wt K} \wh{\wt K}\wt K 
=(\wt G(u,v))^2 -\frac{1}{A} \int\d y\, (\wt G(u,y))^2
-\frac{1}{A} \int\d y\, (\wt G(y,v))^2\\
+\frac{1}{A^2} \int\d x\, \d y\, (\wt G(x,y))^2
+\ldots \ ,
\end{multline}
 \begin{multline}\label{kkkhatkhat}
 \int\d x\, \d y\, \wt K \wt K \wh{\wt K} \wh{\wt K} = 
\wt G(u,v) \frac{1}{4\pi}\left(-\ln\m^2(t_2+t_3) +4\pi \wt G_\zeta(u)-\g\right) \\
-\frac{1}{A} \int\d y\, \wt G(u,y)\wt G(y,v) 
-\frac{1}{A} \int\d y\, \wt G_\zeta(y)\wt G(y,v) +\ldots \ ,
\end{multline}
 \begin{multline}\label{khatkhatkk}
 \int\d x\, \d y\,  \wh{\wt K} \wh{\wt K}\wt K \wt K = 
\wt G(u,v) \frac{1}{4\pi}\left(-\ln\m^2(t_2+t_3) +4\pi \wt G_\zeta(v)-\g\right) \\
-\frac{1}{A} \int\d y\, \wt G(u,y)\wt G(y,v) 
-\frac{1}{A} \int\d y\, \wt G(u,y) \wt G_\zeta(y)+\ldots \ ,
\end{multline}
 \begin{multline}\label{khatkkkhat}
 \int\d x\, \d y\,  \wh{\wt K} \wt K \wt K \wh{\wt K}= 
-\wt G(u,v) \frac{1}{4\pi}\frac{\a_2\a_3}{(\a_2+\a_3)^2} \\
+\frac{1}{4\pi} \left(\frac{1}{t_2+t_3}+\frac{7}{6}R_* - \frac{\a_2\a_3}{(\a_2+\a_3)^2}R_* -\frac{8\pi}{A}\right) \int\d y  \wh{\wt K}(t_1,u,y)\wh{\wt K}(t_4,y,v) +\ldots \ ,
\end{multline}
 \ba\label{khatkhatdtkkhat}
 \int\d x\, \d y\,  \wh{\wt K} \wh{\wt K}\left(-\frac{\d}{\d t} \wt K \right) \wh{\wt K}&=& 
-\wt G(u,v) \frac{1}{4\pi} \left[ \ln\m^2(t_2+t_3)+\g+1 + \frac{\a_2\a_3}{(\a_2+\a_3)^2}\right]
\nonumber \\
&&\hskip-3.cm+\frac{1}{4\pi} \left[\frac{1}{t_2+t_3}- R_*\left(-\frac{1}{6}+\g +\ln\m^2(t_2+t_3)+\frac{\a_2\a_3}{(\a_2+\a_3)^2}\right)-\frac{4\pi}{A}\right] \times
\nonumber \\
&&\hskip-2.cm\times \int\d y\,  \wh{\wt K}(t_1,u,y)\wh{\wt K}(t_4,y,v)
\nonumber \\
&&\hskip-5.cm+\int\d y\, \del_y^i\wt G(u,y)\, \wt G_\zeta(y) \,\del_y^i\wt G(y,v) 
+\frac{R_*}{A}\int\d x \d y\, \wt G(u,x) \wt G(x,y) \wt G(y,v)
+\ldots \ ,
\ea
 \begin{multline}\label{kkhatkhatkhat}
 \int\d x\, \d y\,  \wt K \wh{\wt K} \wh{\wt K}  \wh{\wt K}= 
\int\d y\, (\wt G(u,y))^2\,\wt G(y,v) -\frac{1}{A}\int\d x\d y\, (\wt G(x,y))^2\, \wt G(y,v) +\ldots \ ,
\end{multline}
 \begin{multline}\label{khatkhatkhatk}
 \int\d x\, \d y\,  \wh{\wt K} \wh{\wt K}  \wh{\wt K} \wt K= 
\int\d y\, \wt G(u,y) (\wt G(y,v))^2\,-\frac{1}{A}\int\d x\d y\, \wt G(u,x) (\wt G(x,y))^2\, +\ldots \ ,
\end{multline}
 \begin{multline}\label{khatkkhatkhat}
 \int\d x\, \d y\,  \wh{\wt K} \wt K \wh{\wt K}  \wh{\wt K} = 
\frac{1}{4\pi} \int\d y\, \left[ -\ln\m^2(t_2+t_3)+4\pi\wt G_\zeta(y)-\g \right] \wh{\wt K}(t_1,u,y)\wh{\wt K}(t_4,y,v) \\
-\frac{1}{A}\int\d x\d y\, \wt G(u,x)\, \wt G(x,y)\, \wt G(y,v)\, +\ldots \ ,
\end{multline}
\be\label{khatkhatkhatkhat}
\int\d x\, \d y\,  \wh{\wt K} \wh{\wt K}  \wh{\wt K}  \wh{\wt K} = 
\int\d x\d y\, \wt G(u,x) \,(\wt G(x,y))^2\, \wt G(y,v)\, +\ldots \ . \qquad\qquad\qquad
\ee
Combining everything, we get
\begin{multline}\label{u-O-v2}
\Gd_{u-\hskip-0.8mm{\rm O}\hskip-0.8mm-v}(u,v)
=\frac{1}{2\k^2}\Bigg\{
\frac{1}{\pi}\wt G(u,v) \left[ 3 \ln \frac{\L^2}{\m^2} -3\ln(\a_2+\a_3) -3\g-1-\frac{2\a_2\a_3}{(\a_2+\a_3)^2}\right]\\
+2(\wt G(u,v))^2
-\frac{2}{A}\int\d y (\wt G(u,y))^2 
-\frac{2}{A}\int\d y (\wt G(y,v))^2 
+\frac{2}{A^2}\int\d x \d y (\wt G(x,y))^2 \\
+4 \wt G(u,v)  \big(\wt G_\zeta(u)+\wt G_\zeta(v)\big) 
-\frac{4}{A}\int\d y \big(\wt G(u,y) \wt G_\zeta(y)+\wt G_\zeta(y) \wt G(y,v)\big) \\
+\frac{1}{\pi}\int \d y \wh{\wt K}(t_1,u,y)\wh{\wt K}(t_4,y,v) 
\Big[\frac{2\L^2}{\a_2+\a_3} +R_*\Big( 3\ln\frac{\L^2}{\m^2}-3\ln(\a_2+\a_3) -3\g\\
\hskip2.cm+\frac{4}{3}
-\frac{2\a_2\a_3}{(\a_2+\a_3)^2}\Big) +8\pi R_* \wt G_\zeta(y) -\frac{20\pi}{A}\Big]
\\ 
+4\int\d y\, \del_y^i\wt G(u,y)\, \wt G_\zeta(y) \,\del_y^i\wt G(y,v) \hskip7.cm \\
+2R_*\int\d y\, \left[ (\wt G(u,y))^2\,\wt G(y,v) +\wt G(u,y) (\wt G(y,v))^2\right] 
\hskip4.cm\\
-\frac{4R_*}{A} \int\d x\d y\, \wt G(u,x)\wt G(x,y) \wt G(y,v)\hskip6.5cm\\
-\frac{2R_*}{A}\int\d x\d y\,  (\wt G(x,y))^2 \left[ \wt G(u,x) +\wt G(y,v)\right] \hskip5.cm\\
+2R_*^2 \int\d x\d y\, \wt G(u,x) \,(\wt G(x,y))^2\, \wt G(y,v)\, +\ldots 
\Bigg\} \ .\hskip5.cm
\end{multline}

\def\diaguTOv{\begin{picture}(26,20)(0,0)
\put(0,3.5){\line(1,0){20}}\put(10,15.5){\circle{10}}\put(10,3.5){\line(0,0){7}}
\end{picture}}
\def\diaguTOvsmall{\begin{picture}(10,10)(0,0)
\put(0,1.5){\line(1,0){10}}\put(5,6){\circle{4}}\put(5,1.5){\line(0,0){2.5}}
\end{picture}}
\def\tadpole{\begin{picture}(26,20)(0,0)
\put(10,15.5){\circle{10}}\put(10,5.5){\line(0,0){5}}
\end{picture}}

\subsubsection{One-loop contribution from the tadpole diagram}

The diagram $u\ \diaguTOv v$ only gives finite contributions. We first evaluate the tadpole
\ba\label{tadpole}
-\frac{1}{\sqrt{2}\k} B(x) \equiv \tadpole\hskip-7mm x
= -\frac{1}{\sqrt{2}\k} \int\d y \left[ \wt K \wh{\wt K} + 2 \wh{\wt K} \wt K + R_* \wh{\wt K}\wh{\wt K}\right] \ ,
\ea
where $\wt K \wh{\wt K}\equiv \wt K(t_3,x,y) \wh{\wt K}(t_4,y,y)$ and similarly for the other terms. This simplifies considerably since $\wt K(t_i,y,y)$ does not depend on $y$ and in $\wh{\wt K}(t_j,y,y)$ the only non-constant term is $\wt G_\zeta(y)$. Thus one finds
\ba\label{tadpole2}
B(x)&=&\int\d y\, \wh{\wt K}(t_3,x,y) \D_* \wt G_\zeta(y)\nonumber\\
&=& \wt G_\zeta(x) + R_*\int\d y\,  \wh{\wt K}(t_3,x,y) \wt G_\zeta(y) -\frac{1}{A} \int\d y\, \wt G_\zeta(y) \ .
\ea
The contribution to the Green's function then is
\be\label{GuTOv}
\Gd_{u\, \diaguTOvsmall \,v}(u,v)=\frac{1}{\k^2} \int\d x\, \left(\wt K B \wh{\wt K} + \wh{\wt K} B \wt K + \wh{\wt K} \D_* \wt G_\z \wh{\wt K} + R_* \,\wh{\wt K} B \wh{\wt K} \right) \ ,
\ee
where $\wt K B \wh{\wt K} \equiv \wt K(t_1,u,x) B(x) \wh{\wt K}(t_2,x,v)$ and similarly for the other terms. Inserting the expression for $B$ and evaluating the integrals gives
\begin{multline}\label{GuTOv2}
\Gd_{u\, \diaguTOvsmall \,v}(u,v)=\frac{1}{2\k^2} \Bigg\{
\wt G(u,v) \Big[ 4\wt G_\zeta(u)+4\wt G_\zeta(v) -\frac{4}{A}\int\d x\, \wt G_\zeta(x)\Big]\\
+2R_*\,  \wt G(u,v) \int\d x\, \left(\wt G(u,x)+\wt G(v,x)\right) \wt G_\zeta(x) \\
\hskip2.cm-\frac{2}{A} \int\d x\, \left(\wt G(u,x)+\wt G(v,x)\right) \left( 2\wt G_\zeta(x) +R_*\int\d y\, \wt G(x,y) \wt G_\zeta(y)\right)\\
+6 R_* \int\d x\, \wt G(u,x)\, \wt G_\zeta(x)\, \wt G(x,v) 
-4\int\d x\, \del_x^i\wt G(u,x)\, \wt G_\zeta(x) \,\del_x^i\wt G(x,v) \\
+2R_*^2 \int\d x\d y\, \wt G(u,x)\wt G(x,v)\wt G(x,y) \wt G_\zeta(y) \\
-\frac{2R_*}{A} \int\d x\, \wt G(u,x)\wt G(x,v) \int\d y\, \wt G_\zeta(y)
\Bigg\}
+\ldots\hskip2.cm
\end{multline}

\def\diaguIOv{\begin{picture}(26,20)(0,0)
\put(0,3.5){\line(1,0){20}}\put(10,8.9){\circle{10}}
\end{picture}}
\def\diaguIOvsmall{\begin{picture}(10,10)(0,0)
\put(0,1.5){\line(1,0){10}}\put(5,5){\circle{6}}
\end{picture}}

\subsubsection{One-loop contribution from the quartic vertices}

The diagram $u\ \diaguIOv v$ gets again different contributions:
\ba\label{GuIOv}
\hskip-1.cm\Gd_{u\, \diaguIOvsmall \,v}(u,v)=-\frac{1}{\k^2} \int\d x\, \Bigg\{ \hskip-0.7cm&&
2\wh{\wt K}(t_1,u,x) (\D_*^x-2R_*)\Big( \wh{\wt K}(t_2,z,x)\wh{\wt K}(t_4,x,v)\Big)\Big\vert_{z=x}\nonumber\\
&&+\ \wh{\wt K}(t_1,u,x)\wh{\wt K}(t_4,x,v) (\D_*^x-2R_*) \wh{\wt K}(t_2,x,x)\nonumber\\
&&+\ 2 \big(\D_*^x  \wh{\wt K}(t_1,u,x)\big) \wh{\wt K}(t_2,x,x) \wh{\wt K}(t_4,x,v)\nonumber\\
&&+\ 2 \wh{\wt K}(t_1,u,x) \wh{\wt K}(t_2,x,x) \D_*^x  \wh{\wt K}(t_4,x,v)\nonumber\\
&&+\ 4 \wh{\wt K}(t_1,u,x) \big(\D_*^x \wh{\wt K}(t_2,x,z) \big)\Big\vert_{z=x}  \wh{\wt K}(t_4,x,v)\Bigg\} \ .
\ea
One finds, using the fact that $\wh{\wt K}(t_2,x,x)-\wt G_\zeta(x)$ does not depend on $x$,
\ba
\label{khatdeltakhatkhat}
&&\hskip-1.cm\int\d x\, \wh{\wt K}(t_1,u,x) (\D_*^x-2R_*)\Big( \wh{\wt K}(t_2,z,x)\wh{\wt K}(t_4,x,v)\Big)\Big\vert_{z=x}\nonumber\\
&&= \frac{1}{4\pi}\wt G(u,v)\left( -\ln \m^2 t_2-\g\right) 
+\int\d x\, \del_x^i\wt G(u,x)\, \wt G_\zeta(x) \,\del_y^i\wt G(x,v) \nonumber\\
&&\ \ +\frac{1}{4\pi}\int \d x\, \wh{\wt K}(t_1,u,x)  \wh{\wt K}(t_4,x,v) \left( \frac{1}{t_2} + \frac{7}{6}R_* -\frac{4\pi}{A} -4\pi R_*\wt G_\zeta(x)\right)+\ldots \ , \hskip1.cm
\ea
and
\ba\label{kahtkhatdeltakhat}
&&\hskip-1.cm\int\d x\, \wh{\wt K}(t_1,u,x)\wh{\wt K}(t_4,x,v) (\D_*^x-2R_*)\Big( \wh{\wt K}(t_2,x,x)\Big)\nonumber\\
&&=\wt G(u,v)\left( \wt G_\zeta(u)+\wt G_\zeta(v)\right)
+\frac{1}{4\pi}  \int \d x\, \wh{\wt K}(t_1,u,x)  \wh{\wt K}(t_4,x,v) 2 R_* \left(\ln \m^2 t_2+\g\right)\nonumber\\
&&-2 \int\d x\, \del_x^i\wt G(u,x)\, \wt G_\zeta(x) \,\del_y^i\wt G(x,v) 
-\frac{1}{A} \int\d x\, \left(\wt G(u,x)+\wt G(v,x)\right) \wt G_\zeta(x) \, .\hskip1.cm
\ea
The three remaining terms in \eqref{GuIOv} are straightforward to evaluate:
\ba
&&\hskip-5.mm\int\d x\, \left[\big(\D_*^x  \wh{\wt K}(t_1,u,x)\big) \wh{\wt K}(t_2,x,x) \wh{\wt K}(t_4,x,v)
+\wh{\wt K}(t_1,u,x) \wh{\wt K}(t_2,x,x) \D_*^x  \wh{\wt K}(t_4,x,v)\right]\nonumber\\
&&=\wt G(u,v) \big[\wh{\wt K}(t_2,u,u)+ \wh{\wt K}(t_2,v,v)\big]
-\frac{1}{A} \int\d x\, \wh{\wt K}(t_2,x,x) \big[ \wh{\wt K}(t_1,u,x)+ \wh{\wt K}(t_4,x,v)\big]\nonumber\\
&&\ \ +  \frac{1}{4\pi} \int\d x\, \wh{\wt K}(t_1,u,x)   \wh{\wt K}(t_4,x,v) \,2 R_* \left[  -\ln\m^2 t_2 -\g +4\pi \wt G_\zeta(x) \right] \ ,
\ea
and
\ba
&&\hskip-5.mm\int\d x\, \wh{\wt K}(t_1,u,x) \big(\D_*^x \wh{\wt K}(t_2,x,z) \big)\Big\vert_{z=x}  \wh{\wt K}(t_4,x,v)\nonumber\\
&&\hskip-3.mm= \frac{1}{4\pi}  \int \d x\, \wh{\wt K}(t_1,u,x)  \wh{\wt K}(t_4,x,v) \left(\frac{1}{t_2}+\frac{7}{6} R_* -\frac{4\pi}{A}-R_*\ln\m^2 t_2 -\g R_* +4\pi R_* \wt G_\zeta(x)\right)\ . \nonumber\\
\ea
Combining everything gives
\ba\label{GuIOv2}
\Gd_{u\, \diaguIOvsmall \,v}(u,v)&=&\frac{1}{2\k^2} \Bigg\{
 \wt G(u,v) \left[ -6\wt G_\zeta(u)-6\wt G_\zeta(v) +\frac{1}{\pi}\left(3\ln\m^2 t_2 +3\g\right)\right] \nonumber\\
&&\hskip-3.cm+\frac{1}{\pi}\int \d x \wh{\wt K}(t_1,u,x)\wh{\wt K}(t_4,x,v) \left[ -\frac{3}{t_2} -\frac{7}{2}R_* +3\g R_* -12\pi R_* \wt G_\zeta(x) +3 R_* \ln\m^2 t_2 +\frac{12\pi}{A}\right] 
\nonumber\\
&&\hskip-3.cm +\frac{6}{A} \int\d x\, \left[ \wt G(u,x)+\wt G(v,x)\right] \wt G_\zeta(x) \Bigg\}\ .
\ea

\subsubsection{One-loop contribution from the measure}

Finally the measure vertex \eqref{measurevertex} contributes
\ba\label{G2measure}
\hskip-1.cm \Gd_{\rm measure}(u,v)&=&\frac{1}{\k^2} \int\d x\,  \wh{\wt K}(t_1,u,x) \wt K(t_2,x,x) \wh{\wt K}(t_4,x,v)\nonumber\\
&=&\frac{1}{2\k^2} \frac{1}{\pi}\int\d x\,  \wh{\wt K}(t_1,u,x)  \wh{\wt K}(t_4,x,v) \left[\frac{1}{2t_2} +\frac{7}{12} R_*-\frac{2\pi}{A} \right] \ .
\ea


\end{appendix}

\vskip5.mm
\noindent
{\Large\bf Acknowledgements}

\vskip3.mm
\noindent
A.B. is greatful to Frank Ferrari for the collaboration on \cite{BFunpub}  from which the present project has grown. L.L. is supported by a fellowship from Capital Fund Management.

\vskip5.mm


\end{document}